\documentclass{aa}
\usepackage[varg]{txfonts}

\usepackage{natbib}
\bibpunct{(}{)}{;}{a}{}{,} 

\usepackage{graphicx}

\begin{document} 

   \title{Multipoint study of the rapid filament evolution during a confined C2 flare on 28 March 2022, leading to eruption}
   


\author{Stefan Purkhart\inst{1},
          Astrid M. Veronig\inst{1}\fnmsep\inst{2},
          Bernhard Kliem\inst{3},
          Robert Jarolim\inst{4,1},
          Karin Dissauer\inst{5},
          Ewan C. M. Dickson\inst{1},
          Tatiana Podladchikova\inst{6},
          Säm Krucker\inst{7}\fnmsep\inst{8}
          }

   \institute{Institute of Physics, University of Graz, Universitätsplatz 5, 8010 Graz, Austria\\
             \email{stefan.purkhart@uni-graz.at}
             \and
             Kanzelhöhe Observatory for Solar and Environmental Research, University of Graz, Kanzelhöhe 19, 9521 Treffen, Austria
             \and
             Institute of Physics and Astronomy, University of Potsdam, Potsdam 14476, Germany
             \and
             High Altitude Observatory, 3080 Center Green Dr., Boulder, CO 80301, USA
             \and
             NorthWest Research Associates, 3380 Mitchell Ln, Boulder, CO 80301, USA
             \and
             Skolkovo Institute of Science and Technology, Bolshoy Boulevard 30, bld. 1, Moscow 121205, Russia
             \and
             Institute for Data Science, University of Applied Sciences and Arts Northwestern Switzerland (FHNW), Bahnhofstrasse 6, 5210 Windisch, Switzerland
             \and
             Space Sciences Laboratory, University of California, 7 Gauss Way, 94720 Berkeley, USA
             }

   \date{Received ; accepted }

 
  \abstract
   {}
   {This study focuses on the rapid evolution of the solar filament in active region 12975 during a confined C2 flare on 28 March 2022, which finally led to an eruptive M4 flare 1.5 h later. The event is characterized by the apparent breakup of the filament, the disappearance of its southern half, and the flow of the remaining filament plasma into a new, longer channel with a topology very similar to an extreme ultraviolet (EUV) hot channel observed during the flare. In addition, we outline the emergence of the original filament from a sheared arcade and discuss possible drivers for its rise and eruption.}
   {We took advantage of Solar Orbiter's favorable position, 0.33 AU from the Sun, and $83. 5^\circ$ west of the Sun-Earth line, to perform a multi-point study using the Spectrometer Telescope for Imaging X-rays (STIX) and the Extreme Ultraviolet Imager (EUI) in combination with the Atmospheric Imaging Assembly (AIA) and the Helioseismic and Magnetic Imager (HMI) onboard the Solar Dynamics Observatory (SDO) and H$\alpha$ images from the Earth-based Kanzelhöhe Observatory for Solar and Environmental Research (KSO) and the Global Oscillation Network Group (GONG). While STIX and EUI observed the flare and the filament's rise from close up and at the limb, AIA and HMI observations provided highly complementary on-disk observations from which we derived differential emission measure (DEM) maps and nonlinear force-free (NLFF) magnetic field extrapolations.}
   {According to our pre-flare NLFF extrapolation, field lines corresponding to both filament channels existed in close proximity before the flare. We propose a loop-loop reconnection scenario based on field structures associated with the AIA 1600 Å flare ribbons and kernels. It involves field lines surrounding and passing beneath the shorter filament channel, and field lines closely following the southern part of the longer channel. Reconnection occurs in an essentially vertical current sheet at a polarity inversion line (PIL) below the breakup region, which enables the formation of the flare loop arcade and EUV hot channel. This scenario is supported by concentrated currents and free magnetic energy built up by antiparallel flows along the PIL before the flare, and by non-thermal X-ray emission observed from the reconnection region. The reconnection probably propagated to involve the original filament itself, leading to its breakup and new geometry. This reconnection geometry also provides a general mechanism for the formation of the long filament channel and realizes the concept of tether cutting. It was probably active throughout the filament's continuous rise phase, which lasted from at least 30 min before the C2 flare until the filament eruption. The C2 flare represents a period of fast reconnection during this otherwise more steady period, during which most of the original filament was reconnected and joined the longer channel.}
   {These results demonstrate how rapid changes in solar filament topology can be driven by loop-loop reconnection with nearby field structures, and how this can be part of a long-lasting tether-cutting reconnection process. They also illustrate how a confined precursor flare due to loop-loop reconnection (Type I) can contribute to the evolution towards a full eruption, and that they can produce a flare loop arcade when the contact region between interacting loop systems has a sheet-like geometry similar to a flare current sheet.}

   
   \titlerunning{rapid filament evolution during a C2 flare}
   \authorrunning{Purkhart et al.}

   \maketitle
   

\section{Introduction}\label{sec:introduction}

Solar filaments consist of cold, dense chromospheric plasma that is suspended in the hot corona above photospheric polarity inversion lines (PILs) and are observed as dark structures on the solar disk \citep[e.g., reviews by][]{Labrosse2010,Parenti2014,Gibson2018,Chen2020}.
Their magnetic structure is thought to consist of either sheared arcades with magnetic dips \citep[e.g.,][]{Kippenhahn1957,Antiochos1994,DeVore2000,Aulanier2002,Terradas2015} or twisted magnetic flux ropes \citep[e.g.,][]{Kuperus1974,van_Ballegooijen1989,Low1995,Aulanier1998,Amari1999,Amari2000,Lites2005}. 
Both have U-shaped field line sections that together form the filament channel in which the dense filament plasma can be contained and whose upward magnetic tension force provides support against gravity. However, it has also been shown that magnetic dips are not always necessary for filament formation \citep{Karpen2001}.

Flux ropes can either form below the photosphere and then rise buoyantly into the corona, dragging chromospheric plasma with them \citep[levitation model;][]{Rust1994,Manchester2004,Okamoto2008} or form above the surface from a sheared arcade by twist injection through photospheric motions \citep{Priest1989} or via reconnection \citep{van_Ballegooijen1989}. Chromospheric plasma can also be deposited into a filament channel in the corona by direct injection \citep[injection model;][]{Wang1999,Chae2003} or by evaporation and subsequent condensation due to thermal instability \citep[evaporation-condensation model;][]{Antiochos1999}. Both might also be considered as a unified process \citep{Huang2021}.

Filaments evolve continuously throughout their lifetime, including rapid topology changes due to reconnection. Observations include the interaction of a filament with another nearby filament, sometimes merging into a longer channel \citep[e.g.,][]{Schmieder2004,Joshi2014,Zhu2015,Yang2017,Huang2023,Li2023}, or the reconnection of the filament channel with other nearby magnetic structures \citep{Li2016,Dai2022,Huang2023}.

Their evolution can eventually end in a filament eruption, which is associated with major space weather events such as flares and coronal mass ejections \citep[CME;][]{Parenti2014}. There are several models for their destabilization and eruption, which can be divided into resistive processes such as tether cutting \citep{Moore1992,Moore2001}, breakout reconnection \citep{Antiochos1999,Lynch2008,Karpen2012}, flux cancellation \citep{van_Ballegooijen1989}, and flux emergence \citep{ChenShibata2000}, and ideal magnetohydrodynamic instabilities such as the torus instability \citep{Kliem2006} and the kink instability \citep{Hood1981,Torok2004}.

Sometimes filaments undergo a failed eruption where their ejection stops after a certain height \citep[e.g.,][]{Ji2003}, which may be caused by strong overlying magnetic fields \citep{Torok2005} or by reconnection of the sheared field lines with the overlying fields \citep{DeVore2008}. In a few cases, the filament eruption has been observed to successfully resume at a later time, resulting in a two-stage eruption \citep{Byrne2014,Gosain2016,Chandra2017}.

The filament under study erupted on 28 March 2022, in active region (AR) 12975, in association with an M4 flare that was extensively analyzed by \citet{Purkhart2023}. A relevant finding was a non-thermal hard X-ray (HXR) source that jumped from a flare ribbon to the southern anchor point of the erupting filament during a late HXR peak. This followed an otherwise continuous westward drift throughout the flare, as expected from asymmetric filament eruption models \citep{Liu2009}, in which reconnection propagates along a series of overlying arcades. Their nonlinear force-free (NLFF) magnetic field extrapolations extended this model by showing that the early phase of reconnection involved strongly sheared magnetic field lines close to the filament channel, which were probably critical for the later evolution of the HXR source.

In this paper we study the evolution of the filament before its eruption. In particular, we focus on its rapid restructuring during a confined C2 flare at 09:56 UT on 28 March 2022. Observations show an apparent breakup of the original short filament during the flare, the disappearance of its southern half, and the flow of the remaining filament plasma into a longer channel with a different southern footpoint. This new longer filament erupted about one hour later. In addition, we provide an overview of the AR emergence, the original filament formation and its first appearance about seven hours before the eruption, and its further evolution up to the studied restructuring during the C2 flare. Combined with the study by \citet{Purkhart2023}, this study gives a complete picture of the evolution of the filament over its lifetime.

\begin{figure}
  \resizebox{\hsize}{!}{\includegraphics{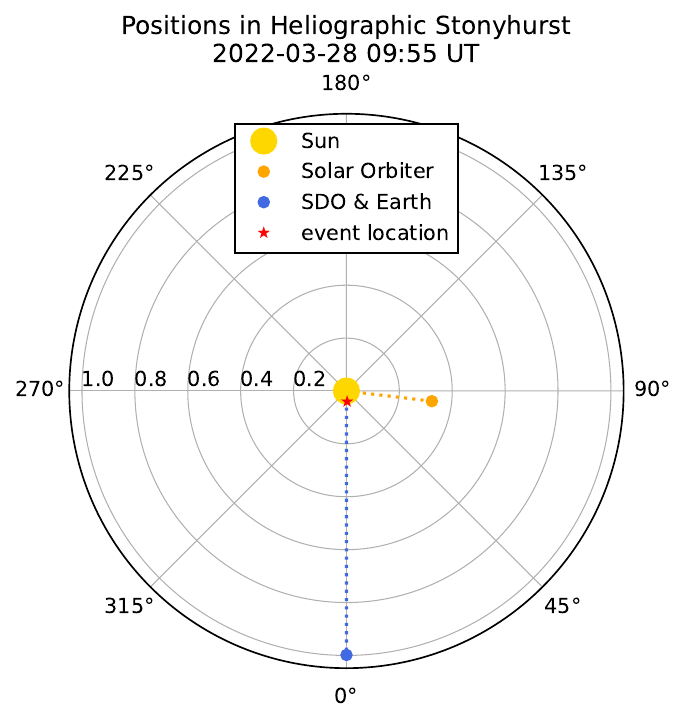}}
  \caption{Positions of the Solar Orbiter, SDO, and Earth in Heliographic Stonyhurst coordinates at the time of the C2 flare being studied (09:55~UT). The location of the event is marked by the red star.}
  \label{fig:SC_positions}
\end{figure}

We take advantage of the multi-point configuration of Solar Orbiter and the Solar Dynamics Observatory (SDO) during the event under study, which is illustrated in Fig. \ref{fig:SC_positions}. Solar Orbiter was near its first science perihelion, at a distance of 0.33~AU from the Sun and a longitudinal separation of $83.5^\circ$ west of the Sun-Earth line. From this perspective, AR 12975 was positioned near the eastern limb, allowing instruments such as the Spectrometer Telescope for Imaging X-rays (STIX) and the Extreme Ultraviolet Imager (EUI) to obtain close-up side-on views of the filament's evolution, rise, and destabilization. The Atmospheric Imaging Assembly (AIA) and the Helioseismic and Magnetic Imager (HMI) onboard SDO observed the AR near the disk center, providing a more complete view of the filament restructuring and flare as well as ideal conditions for the NLFF magnetic field extrapolations. Those observations were further supplemented by H$\alpha$ images by the Kanzelhöhe Observatory for Solar and Environmental Research (KSO) and the Global Oscillation Network Group (GONG).

We analyze observations of the filament restructuring from both the Earth and Solar Orbiter perspectives (Sect.~\ref{sec:results_filament_observations}), examine the role of the C2 flare in this process (Sect.~\ref{sec:results_flare}), and attempt to explain the reconnection using NLFF extrapolations (Sect.~\ref{sec:results_NLFF}). The overview of the AR emergence and original filament's formation is given in Sect.~\ref{sec:results_emergence}.

\section{Data and methods}\label{sec:methods}

\subsection{Solar Orbiter/STIX}

The STIX instrument \citep{Krucker2020_STIX} onboard Solar Orbiter is a HXR spectrometer and imager observing in the 4 to 150 keV energy band. This range is essential for observing the bremsstrahlung emitted by high-energy non-thermal electrons as they are decelerated by the field of ambient ions. In addition, STIX observes thermal emission from the hottest flaring plasma above 10 MK, providing us with plasma diagnostics beyond the temperature response of extreme ultraviolet (EUV) imagers such as AIA and EUI.

The STIX website\footnote{\url{https://datacenter.stix.i4ds.net}} serves as a user-friendly interface to the comprehensive STIX data platform \citep{Xiao2023} that allows users to query, analyze, and download STIX data. For all STIX analyses, we used pixel data, which contain individual counts for each pixel on all detectors. These data are needed for image reconstruction \citep{Massa2023} and also provides great flexibility by allowing us to use only certain pixels for our analysis. A background observation was subtracted from the pixel data to isolate counts due to the observed event, and only the bottom row of pixels was used to account for a partial occultation of the STIX detectors caused by the flare's position relative to the spacecraft pointing at that time. The data were adjusted for the difference in light travel time (335.7 s) from the Sun to Solar Orbiter compared to an Earth-based observation. All times given in this paper are therefore for an Earth-based observer to allow direct comparisons with observations from SDO or KSO. 

The count spectra measured by STIX were assumed to have been generated either by a purely thermal electron distribution or by a combination of thermal and non-thermal electron distributions. We fitted models for both scenarios using the OSPEX\footnote{\url{http://hesperia.gsfc.nasa.gov/ssw/packages/spex/doc/}} software distributed as part of the SolarSoftWare environment for the IDL programming language (SSWIDL). The thermal part of the spectrum was fitted with the isothermal model 'vth', while the higher energy non-thermal spectrum was fitted with the thick target model 'thick2', restricted to a single power law.

STIX imaging was performed using the Clean method \citep[stx\_vis\_clean,][]{Hogbom1974} available in the SSWIDL imaging pipeline. The STIX aspect system \citep{Warmuth2020} optically measures the orientation of STIX relative to the Sun and provides a corrective shift for the source positions. However, for the event under study, the derived aspect solution was not accurate enough and an additional constant shift had to be applied to all STIX images. These corrections were taken from the previous study by \citet{Purkhart2023}, who aligned the non-thermal footpoints observed by STIX with AIA 1600~Å ribbons at the time of the M4 flare associated with the filament eruption.

Comparing STIX images with observations from other perspectives is not always straightforward. HXR sources that do not correspond to flare footpoints, but instead lie within the corona, cannot simply be reprojected to the AIA perspective, since their exact position along the line of sight is unknown. For such cases we instead construct a line from Solar Orbiter through the maximum of an HXR source and then reproject this line onto the AIA images. This technique gives a good indication of which AIA EUV features may correspond to the observed STIX HXR emission and was used to produce results presented in Sect.~\ref{sec:results_flare}. All reprojections and visualizations in this paper used version 5.0.0 \citep{SunPy_5.0.0} of the SunPy open source software package \citep{sunpy_community2020} for the Python programming language.

\subsection{Solar Orbiter/EUI}

For EUV observations, which match the STIX perspective, we used data from EUI \citep{Rochus2020_EUI} onboard Solar Orbiter. Of the three telescopes, only the Full Sun Imager (FSI) provided suitable data for this event because the High Resolution Imagers (HRI) were focused on a different region of the Sun. The FSI observes a field of view of $3.8^\circ \times 3.8^\circ$ in the 174 and 304~Å pass bands, allowing it to produce images about four solar radii wide during Solar Orbiter perihelia. Its angular resolution of 10 arcsec corresponds to a linear resolution of about 2.4~Mm during the event under study. Combined with the limited cadence of 10 min and easily saturated images due to the long and constant exposure time (10~s), the observations were not ideal for flare observations. However, they were very advantageous for following the height evolution of the filament, since the AR was close to the limb from the perspective of Solar Orbiter. Since this event, the usability of FSI data for flare studies has been greatly improved, with the instrument now routinely taking low (0.2~s) exposures in addition to every normal exposure since November 2022. We have downloaded\footnote{\url{https://www.sidc.be/EUI/data/}} EUI L2 images that are part of the Solar Orbiter EUI Data Release 6.0 2023-01 \citep{EUI_datarelease6.0} and corrected the observation times for the difference in light travel time as with the STIX data.

\subsection{SDO/AIA and HMI}\label{sec:methods_AIA_HMI}

For EUV observations with higher resolution, a complementary on-disk view of the event, and the possibility of more detailed temperature diagnostics, we used data from AIA \citep{Lemen2012_AIA} onboard SDO \citep[][]{Pesnell2012_SDO}. AIA also provides full disk images of the Sun, with a spatial resolution of 1.5~arcsec at a pixel scale of 0.6~arcsec, a high cadence of 12 s, and in several different wavelengths.

We used the following six coronal EUV wavelength channels centered on lines produced by the specified iron ions at different peak formation temperatures (log(T[K])): 94~Å (Fe\,{\sc xviii}; 6.8), 131~Å (Fe\,{\sc viii} and Fe\,{\sc xxi}; 5.6 and 7.0), 171~Å (Fe\,{\sc ix}; 5.8), 193~Å (Fe\,{\sc xii} and Fe\,{\sc xxiv}; 6.2 and 7.3), 211~Å (Fe\,{\sc xiv}; 6.3) and 335~Å (Fe\,{\sc xvi}; 6.4). Each wavelength channel is sensitive to a much broader range of temperatures, and together they cover a range of at least $10^5$ to $10^7$ K \citep{Lemen2012_AIA,Boerner2012}, allowing us to reconstruct the emission measures and temperature using DEM analysis.

In addition, we used images from the AIA 304~Å EUV and AIA 1600~Å UV channels. The AIA 304~Å filter is centered at a He\,{\sc ii} line, sensitive to temperatures from the chromosphere to the transition region, and reveals the dynamics of the filament plasma. AIA1600~Å sees C\,{\sc iv} and continuum emission from the upper photosphere and transition region, showing the flare ribbons that typically correspond to the location where the energy of high-energy electrons is deposited.

For DEM analysis (see Sect.~\ref{sec:methods_DEM}), we downloaded full disk Level 1 images from JSOC\footnote{\url{http://jsoc.stanford.edu}} and deconvolved them using the default settings of the $aia\_deconvolve\_richardsonlucy$ SSWIDL function and providing the point spread function obtained from $aia\_calc\_psf$. Standard SSWIDL procedures were used to perform further image calibration for Level 1.5 data, as well as differential rotation compensation and preparation of the final submaps passed to the DEM code.

Those AIA images that were only used for visualization purposes in figures and movies were downloaded as tracked submaps using the cutout feature available from JSOC. The same was done for the line-of-sight magnetograms and continuum images from \citep[][]{Schou2012_HMI} HMI. SunPy was used for all visualizations and reprojections of the SDO data. Finally, we used HMI vector magnetic field data for the NLFF magnetic field extrapolation (see Sect.~\ref{sec:methods_NLFF}).

\subsection{Differential emission measure analysis}\label{sec:methods_DEM}

The solar corona is an optically thin plasma. Assuming that it is in thermal equilibrium, the intensities $I_\lambda$ measured in each wavelength channel $\lambda$ of our instrument can be described by the temperature dependent response function $K_\lambda(T)$ of that filter and by the differential emission measure $DEM(T)$ as

\begin{equation}
    I_\lambda = \int_T K_\lambda(T) DEM(T) dT.
\end{equation}\\
The DEM is typically defined as a function of the electron density $n$ along the line-of-sight $h$ as

\begin{equation}
    DEM(T) = n(h(T))^2dh/dT.
\end{equation}\\
The DEM is therefore a measure of the amount of plasma emitting at a given temperature along a certain line of sight. Calculating the DEM from a set of intensities $I_\lambda$ observed by an instrument is not a straightforward process, since it is convolved by the instrument response and the emission process. 

We used the regularized inversion algorithm developed by \citet{Hannah2012} in its IDL implementation. The prepared AIA submaps (see Sect.~\ref{sec:methods_AIA_HMI}) were further binned by 2x2 pixels to increase the signal-to-noise ratio. We calculated the DEM for 40 equally spaced logarithmic temperature bins between 0.5 and 31.6~MK using the AIA temperature response returned by the \textit{aia\_get\_response} SSWIDL routine using coronal abundances from the CHIANTI 10.0 database \citep{Dere1997_CHIANTI_1,Del_Zanna2021_CHIANTI_10}.

\subsection{Nonlinear force-free magnetic field extrapolation}
\label{sec:methods_NLFF}

The NLFF magnetic field extrapolations were performed for cylindrical equal area (CEA) projections of the entire Spaceweather HMI Active Region Patch \citep[SHARP,][]{Bobra2014} 8088, that included the AR 12975 under study as well as AR 12976 to the east. We used the method developed by \citet{Jarolim2023}, which utilizes a physics-informed neural network to derive a solution for the coronal magnetic field that satisfies the force-free assumption and the photospheric vector magnetic field observations. The method is available via an interactive Google Colab linked in the projects Github\footnote{\url{https://github.com/RobertJaro/NF2}}, which makes it possible to easily run the magnetic field extrapolations from a web browser.
 
The neural network acts as a mesh-free representation of the simulation volume, by mapping coordinate points to the respective magnetic field vectors. The method uses an iterative optimization approach to approximate the boundary condition (i.e., the photospheric vector magnetogram) and to minimize the residuals of the divergence-free and force-free equation in the simulation volume.

From the resulting magnetic field solution we compute the free magnetic energy, which is given by the difference in magnetic energy $E(\vec{B}) = \int_V \frac{\vec{B}^2}{8 \pi} dV$ between the force-free solution and a potential field extrapolation (i.e., $E_{free} = E_{FF} - E_{PF}$). Here, $V$ refers to the simulation volume, $E_{FF}$ to the magnetic energy of the force-free solution, and $E_{PF}$ to the magnetic energy of the potential field solution.

To evaluate the quality of the NLFF extrapolations, we consider three standard metrics that estimate the divergence- and force-freeness of the magnetic field solution \citep[c.f.,][]{Schrijver2006}. We compute the average normalized divergence
\begin{equation}
    L_{div,n}(\vec{B}) = \frac{1}{N} \sum_i^N |\vec{\nabla} \cdot \vec{B}_{i}| / \lVert \vec{B}_{i} \rVert \,,
\end{equation}
where $B_i$ refers to the magnetic field vector and $N$ to the total number of grid cells. To quantify the force-freeness, we compute the current-weighted average sine of the angle between the magnetic field and the electric current density vector (\vec{J} = $\nabla \times \vec{B}$)
\begin{equation}
    \sigma_J(\vec{B}) = \frac{\sum_i \frac{\lVert \vec{J}_i \times \vec{B}_{i} \rVert}{\lVert \vec{B}_{i} \rVert}}{\sum_i \lVert \vec{J}_i \rVert \,}
\end{equation}
and the corresponding angle
\begin{equation}
	\theta_J(\vec{B}) = sin^{-1}(\sigma_J).
\end{equation}\\
Note that force-free extrapolations can only provide an estimate of the coronal magnetic field and are intrinsically limited by neglecting plasma contributions at photospheric and chromospheric heights. This could be improved, for example, by additional chromospheric observations \citep{Fleishman2019,Jarolim2024}, which are not provided on a regular basis.

We used the open source software ParaView\footnote{\url{https://www.paraview.org}} to visualize the magnetic field lines in the NLFF extrapolation from the pre-flare time of 09:36 UT. The background for the presented visualizations was created by overlaying the 09:36~UT HMI line of sight (LOS) magnetogram in CEA projection with the contours of the AIA 1600 Å image at the peak of the flare (09:55:47UT), adjusted for differential rotation. The background image, and in particular the AIA 1600~Å contours, were used to guide the placement of the starting points for the relevant field lines.

\subsection{Kanzelhöhe and GONG H$\alpha$ filtergrams}

Full-disk H$\alpha$ filtergrams were used to follow the evolution and formation of the filament. The data used were mainly from the H$\alpha$ telescope at KSO \citep{Poetzi2021_KSO} in Austria. The refractor is equipped with a Zeiss-Lyot filter with a central wavelength of 6562.8~Å and a FWHM of 0.7~Å and a 2048 x 2048 pixel CCD camera that takes an image about every 6~s. In order to follow the AR emergence and filament formation over a full day, we supplemented the KSO data with H$\alpha$ images taken by the GONG \citep{Harvey1996} network for the time when KSO data were not available.

\section{Results}\label{sec:results}

\subsection{Filament restructuring during a confined C2 flare}\label{sec:results_filament_observations}

\begin{figure*}
   \includegraphics[width=18cm]{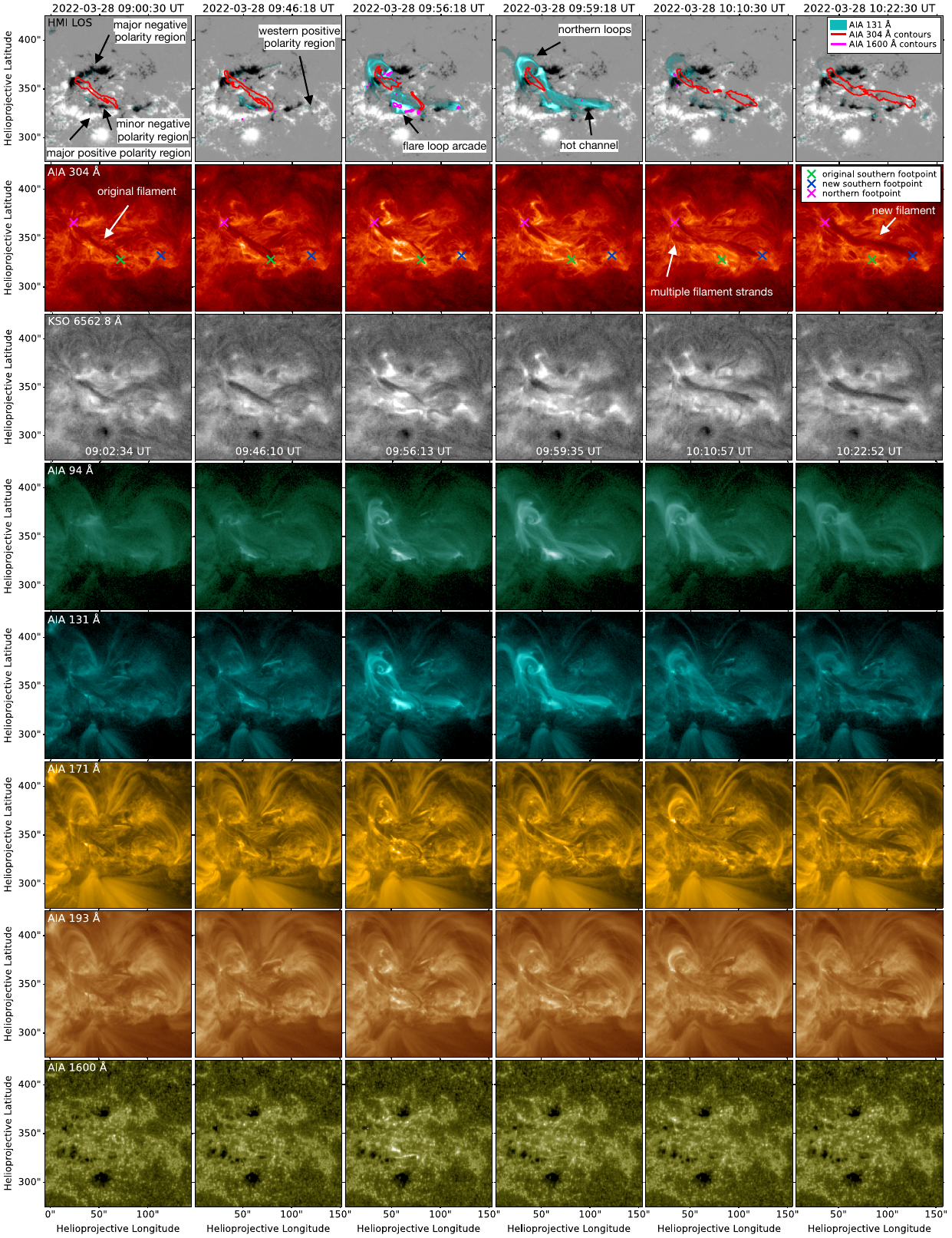}
    \caption{Evolution of the filament as observed by selected SDO/AIA channels and by KSO in H$\alpha$ at six times (columns). Top row: AIA 131~Å images, filament contours extracted from the AIA 304~Å channel, and AIA 1600~Å contours overlayed on HMI LOS magnetograms, scaled to $\pm 1000$~G. The positions of the filament footpoints are marked on the AIA 304~Å images. The associated movie is available online.
    }
    \label{fig:AIA_Cflare_overview}
\end{figure*}

\begin{figure*}
\centering
  \includegraphics[width=18cm]{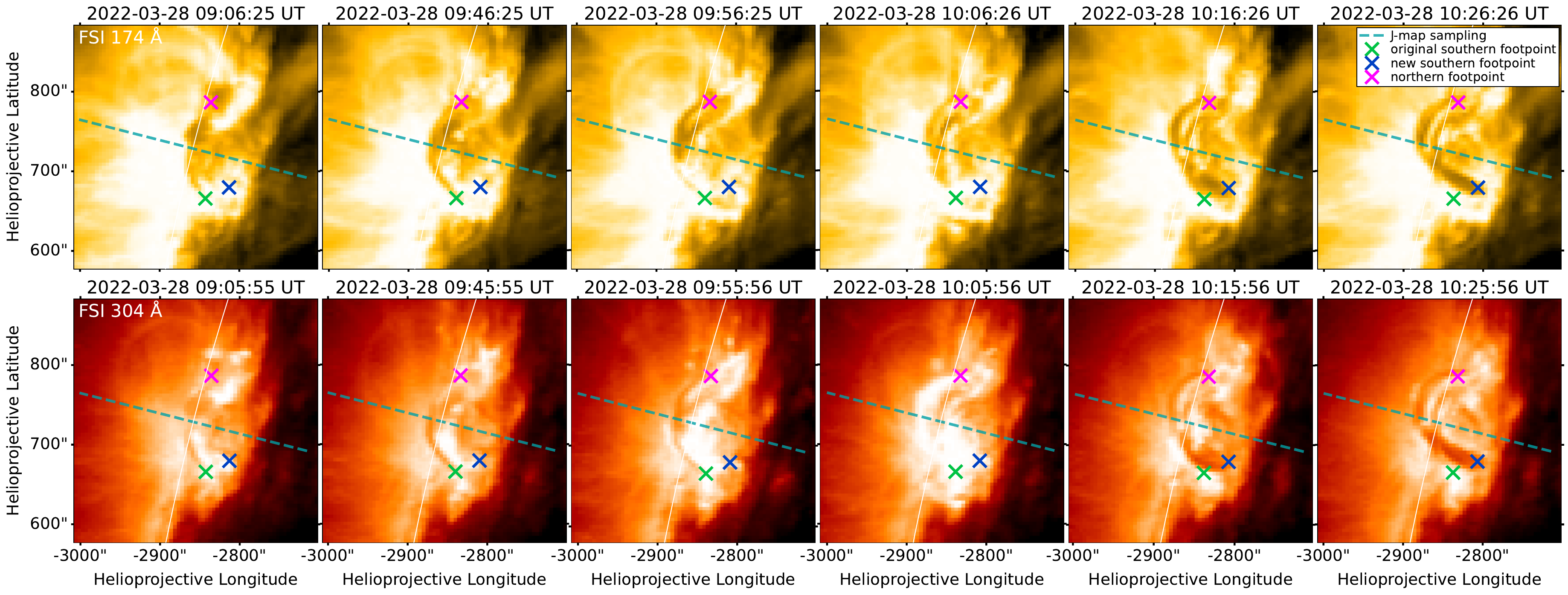}
    \caption{Evolution of the filament as observed by EUI/FSI in the 174 and 304~Å channels at six selected times (columns). The original and new southern filament footpoints are reprojected from AIA (see Fig.~\ref{fig:AIA_Cflare_overview}) assuming a height of 4~Mm above the photosphere. The solar limb is marked by a white line. Intensity profiles sampled along the dashed line (J-map sampling) were used to create Fig. \ref{fig:EUI_Jmap}. The associated movie is available online.}
    \label{fig:EUI_Cflare_overview}
\end{figure*}

\begin{figure}
  \resizebox{\hsize}{!}{\includegraphics{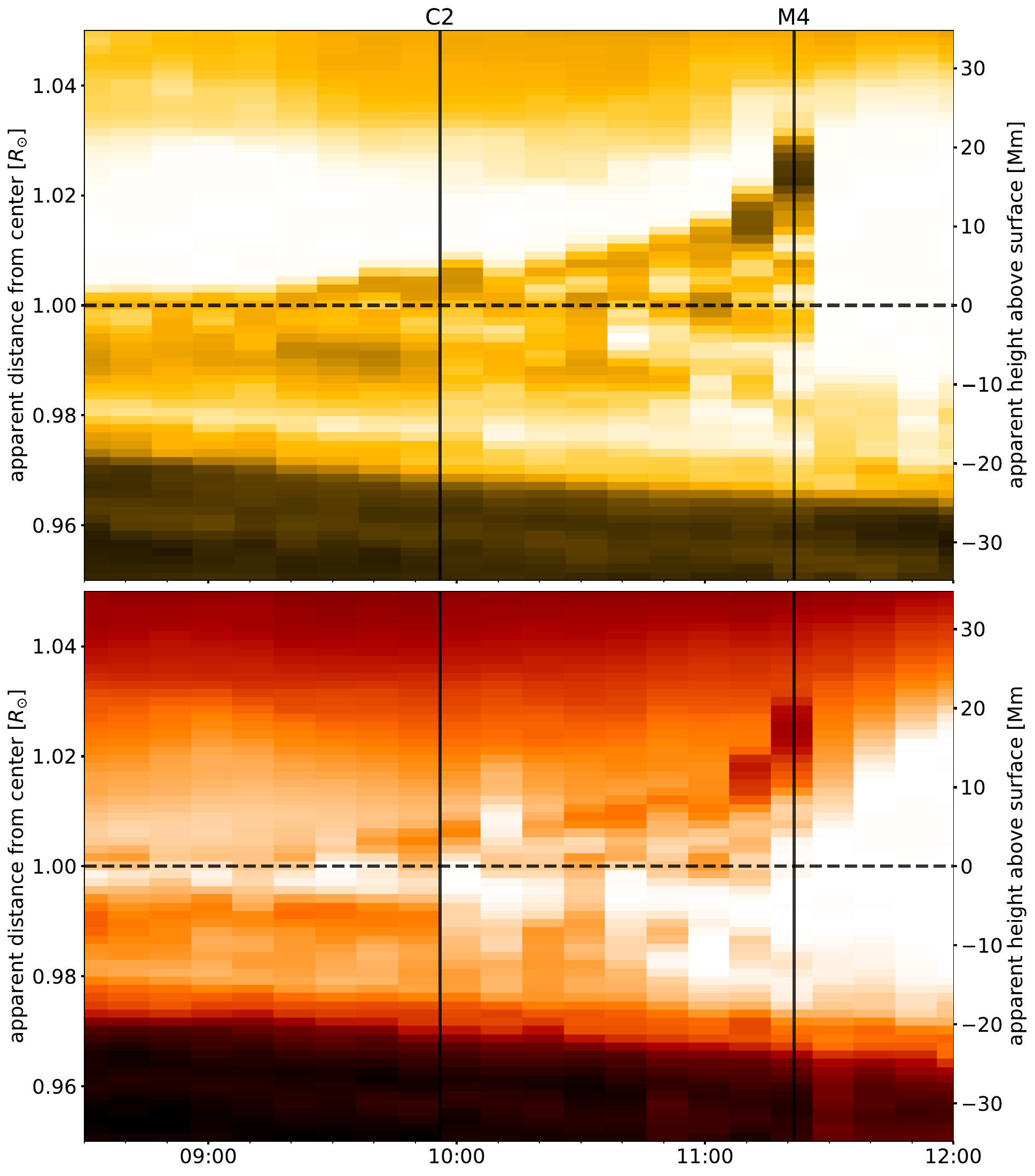}}
  \caption{Evolution of the intensity profile sampled along the marked line in Fig.~\ref{fig:EUI_Cflare_overview}. The times of the peak 15--25~keV HXR flux during the C2 flare and of the first major HXR burst during the M4 flare are marked.}
  \label{fig:EUI_Jmap}
\end{figure}

This section focuses on the rapid evolution of the filament during a confined C2-class flare on 28 March 2022 (GOES start time: 09:48, peak time: 09:57~UT) as observed by SDO/AIA and HMI, KSO, and Solar Orbiter/EUI. During the flare, we observe an apparent breakup of the original filament and the disappearance of its southern part. Subsequently, the remaining plasma appears to flow into another, longer filament channel, resulting in the longer filament that erupted shortly thereafter at about 11:20~UT.


Figure \ref{fig:AIA_Cflare_overview} and its accompanying movie show this evolution of the filament over about 1.5 hours as observed on-disk by AIA (E)UV and KSO H$\alpha$ images. Around 09:00:30~UT the filament can be seen in its original configuration. Comparing the contours of the filament in the AIA 304~Å channel with the HMI LOS magnetogram from that time (top panel), we see that the filament extends from the eastern end of the major negative polarity region in the north of the AR to the western end of the major positive polarity region in the south.

In the period leading up to the C2 flare, we observed several brightenings below the filament which may be related to its destabilization. We show the strongest brightening in AIA 94 and 131 Å at 09:46:18~UT, located between the major positive polarity region and the adjacent minor negative polarity region, which appears to be directly followed by the very gradual onset of the flare. Over the next 10 minutes, this area becomes progressively brighter as a flare loop arcade develops between both polarities from about 09:54~UT. This flare arcade reaches maximum brightness in the AIA 131~Å channel at about 09:56:18~UT. 

An EUV hot channel brightens in AIA 131~Å from about 09:50~UT. Initially, it has footpoints close to the original filament footpoints and shows indications of a right-handed twist around the northern part of the filament. The channel moves rapidly towards the southern part of the filament, causing a diffuse brightening in AIA 131~Å below the filament at about 09:53~UT. From about 09:54 UT the southern part of the channel brightens along a new path that ends in a positive polarity region west of the AR, where the new filament channel also ends. The original southern part of the channel becomes invisible. The new hot channel wraps around the filament, passing just below the filament where it breaks up. 

The southern part of the filament disappears stepwise (thread by thread) in all AIA channels during about 09:52 to 10:01~UT (best seen in AIA 304~Å and KSO H$\alpha$). This is accompanied by irregular brightenings primarily at the underside of the filament and along the disappearing threads. The brightenings eventually concentrate at the edge of the remaining northern part of the filament. 

AIA 1600~Å and KSO H$\alpha$ show the two flare ribbons in the major positive and minor negative polarity regions, mapping to the small flare loop arcade observed in AIA 131~Å. Additional emission kernels are located in the western positive polarity, which corresponds to the footpoint of the extended channel, and in the major negative polarity. 

The original filament’s breakup was largely complete by 09:58~UT. The included images from about 09:59:18~UT show the filament after it has been nearly completely separated from its southern footpoint, with the northern part of the filament still holding dark plasma. The hot channel has brightened further in AIA 131~Å and is now also visible in AIA 94~Å, and appears to extend through the filament breakup region.

Immediately after the complete breakup of the original filament, the remaining filament plasma begins to flow into a new, longer channel, while the emission from the flaring loops and the hot channel fade rapidly. At 10:10:30~UT we see a snapshot of the partially filled new filament channel.

The images taken around 10:22:30~UT show the newly formed filament channel after it is completely filled with dark plasma. A comparison of the filament contours in AIA 304~Å with the HMI LOS magnetogram shows that the filament now extends towards the positive polarity region in the west of the AR, roughly cospatial to the hot channel that formed during the flare. The positions of the original and new southern filament footpoints are marked in the 304~Å images for comparison and have an angular separation of about 48~arcsec, corresponding to about 34~Mm. Their distance from the marked northern footpoint is 61~arcsec (44~Mm) and 101~arcsec (73~Mm), respectively. The apparent position of the northern footpoint has also changed slightly, which could be due to projection effects if the filament has risen during this restructuring. In addition, the filament appears to be more complex in the north, with multiple strands visible throughout the evolution. 

The C2 flare is followed by a quieter phase characterized by sporadic brightenings below the filament channel and even a temporary retraction of the filament plasma from the southern footpoint (see movie). Around 11:00~UT a continuous brightening and further growth of the small loop arcade sets in as the filament destabilizes, leading to its eventual eruption associated with an M4 flare at about 11:20~UT \citep{Purkhart2023}.


Figure \ref{fig:EUI_Cflare_overview} and its accompanying movie show the restructuring of the filament as observed on the eastern solar limb from the Solar Orbiter perspective by EUI/FSI. The evolution is shown in both the 174 and 304~Å filters and covers a similar time frame as the AIA observations in Fig.~\ref{fig:AIA_Cflare_overview}. However, the EUI/FSI time cadence was limited to 10~min during this event, which predetermines the selection of snapshots shown. The marked coordinates for both the original and the new southern filament footpoints were reprojected from the AIA perspective assuming a height of 4~Mm above the photosphere to match the EUI observations.

At 09:06~UT the filament is almost indistinguishable from other features in the AR due to its low height, but half of its curved outline can be seen south of the dashed line.
By 09:46~UT it has risen significantly and is now more clearly visible as a full arch-like structure extending between the northern and the original southern footpoint.
The images taken around 09:56~UT capture the peak of the flare, which can be seen as a brightening on and below the southern part of the filament. The remaining connection of the filament to its southern footpoint appears very thin, matching the observations of AIA and KSO from that time. The complete separation occurred a few minutes later and was unfortunately missed by EUI. The northern part of the filament now appears to be separated into two strands that merge near the top, which may correspond to the separate filament strands observed by AIA.
At 10:06~UT we see that the plasma in the 174~Å channel has started to flow back, consistent with the AIA and H$\alpha$ observations. However, in the 304~Å channel, the flaring region appears even more saturated than during the flare, and a bright loop has formed following the shape of the filament.
By 10:16~UT the dark plasma has mostly filled the new filament channel, and by 11:26~UT we see the fully formed new filament. Even from this perspective, the change in position of the southern footpoint is clearly visible. The northern part of the new filament still appears to consist of two separate strands running mostly parallel to each other. However, the bright gap between them has become much longer compared to 09:56~UT and could indicate some rotation of the filament structure. The side-on view of EUI/FSI also confirms that the filament has risen significantly during the observation period.

In Fig.~\ref{fig:EUI_Jmap} we plot the time evolution of the intensity profiles sampled from both EUI/FSI channels along the dashed line drawn in Fig.~\ref{fig:EUI_Cflare_overview}. The sampling line was chosen to be just south of a dark stationary structure located on the limb from the EUI perspective. This ensures that we best capture the initial motion of the filament, which is otherwise hidden behind this structure, while also positioning the line close to the region of filament breakup during the restructuring. After the C2 flare, the line may appear to be off-center, but it is again consistent with the direction of the filament's motion during the ejection.

Both panels show the apparent height evolution of the filament from 08:30 to 12:00~UT, with the times of the C2 and M4 flares marked. The solar rotation affects these results, most notably seen in the downward drift of the sharp transition from the dark, quiet Sun to the bright AR in the lower half of each panel, which shows an apparent drop in height below the limb of around 5~Mm. This trend makes it difficult to derive absolute heights for the filament and would lead to an underestimation of the rate of rise because it is convolved with the downward trend. Nevertheless, the figure gives a good idea of the general behavior of the filament's height evolution.

We observe a mostly steady rise with an apparent height increase of about 5~Mm from about 09:15 UT until the C2 flare. However, it is unclear if the filament began to rise at this time or if the filament was previously hidden behind other structures in the AR. During the C2 flare, there appears to be a more abrupt rise in height following the slight plateau before the flare. Immediately afterward, at about 10:06 UT, we see a sharp drop in height back to the pre-flare level as the plasma moves into the new filament channel. Comparing this with the full EUI images in Fig.~\ref{fig:EUI_Cflare_overview}, we can see that the height decrease is only observed near the flaring and filament restructuring region, and that the northern part of the filament remains at about the same height during this time (compare images at 09:56 and 10:16 UT). The new filament then immediately resumes its steady, mostly linear rise from this lower height up to an apparent height of about 10~Mm above the limb, with the onset of an exponential takeoff noticeable around 11:00~UT as it finally accelerates towards its eruption accompanied by the M4 flare. This height profile after the C2 flare is consistent with the timing of a quiet phase and the onset of continuous brightening of the small loop arcade seen in the AIA observations and described earlier in this section.

\subsection{The C2 flare}\label{sec:results_flare}

\begin{figure*}
\centering
  \includegraphics[width=18cm]{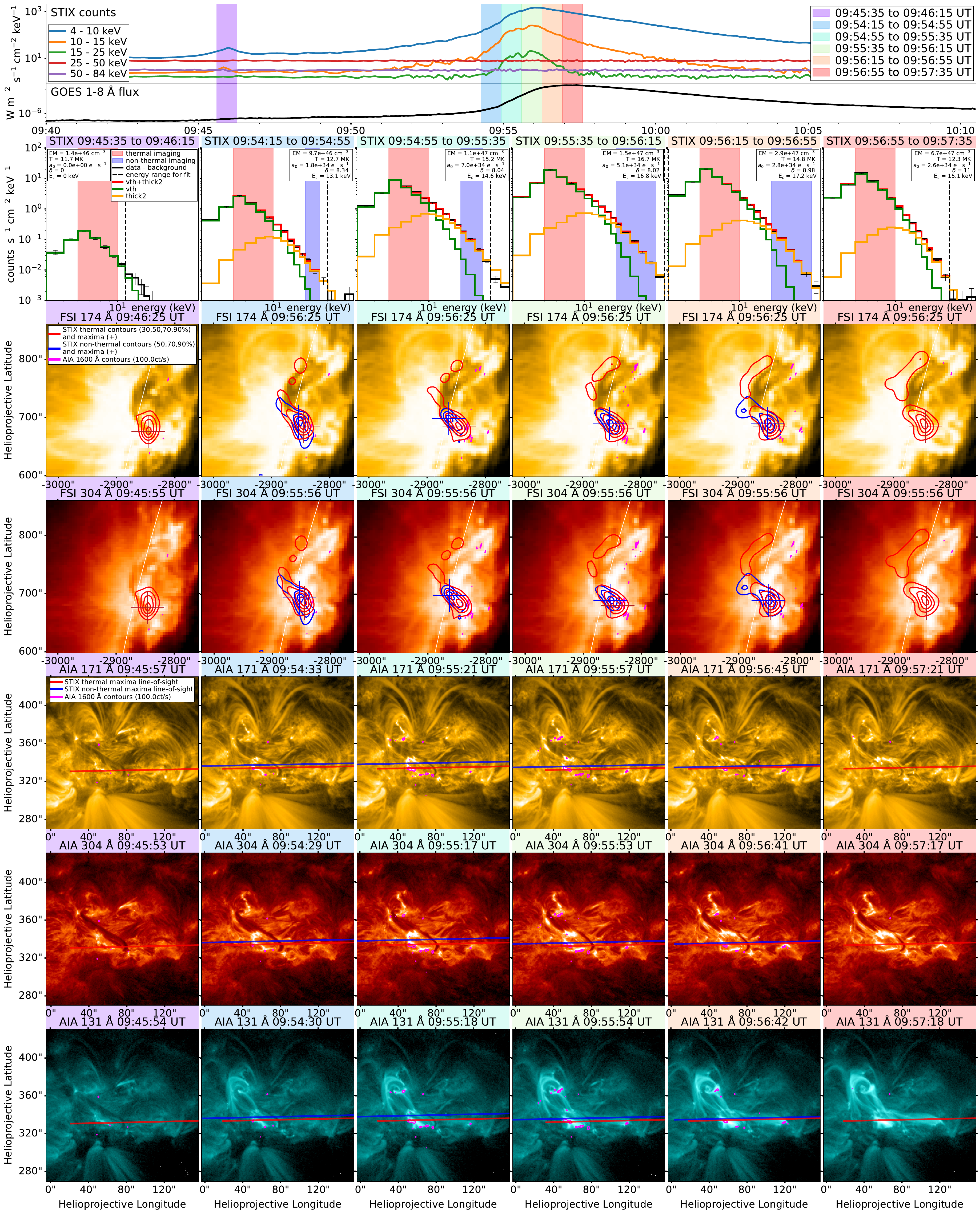}
    \caption{
    Overview of STIX observations and their relation to EUI and AIA images.
    Top row: STIX light curves for five energy ranges from 4 to 84~keV. Imaging intervals are indicated by different colors. 
    Columns: Color-coded according to the marked time intervals containing the following observations:
    1) STIX X-ray spectrum fitted with a thermal (vth) and nonthermal (thick2) electron model. The energy intervals used for imaging are marked. 
    2) STIX clean image contours overlaid on the nearest available EUI FSI 174~Å image. The maxima of the HXR sources from each energy range are marked.
    3) Same as 2) but with the EUI FSI 304~Å image. 
    4) AIA 171~Å image with reprojected lines through the STIX HXR maxima marked in 2) and 3). 
    5) and 6) Same as 4), but with AIA 304 and 131~Å images, respectively.
    }
    \label{fig:STIX_overview_multiple}
\end{figure*}

\begin{figure*}
\centering
  \includegraphics[width=18cm]{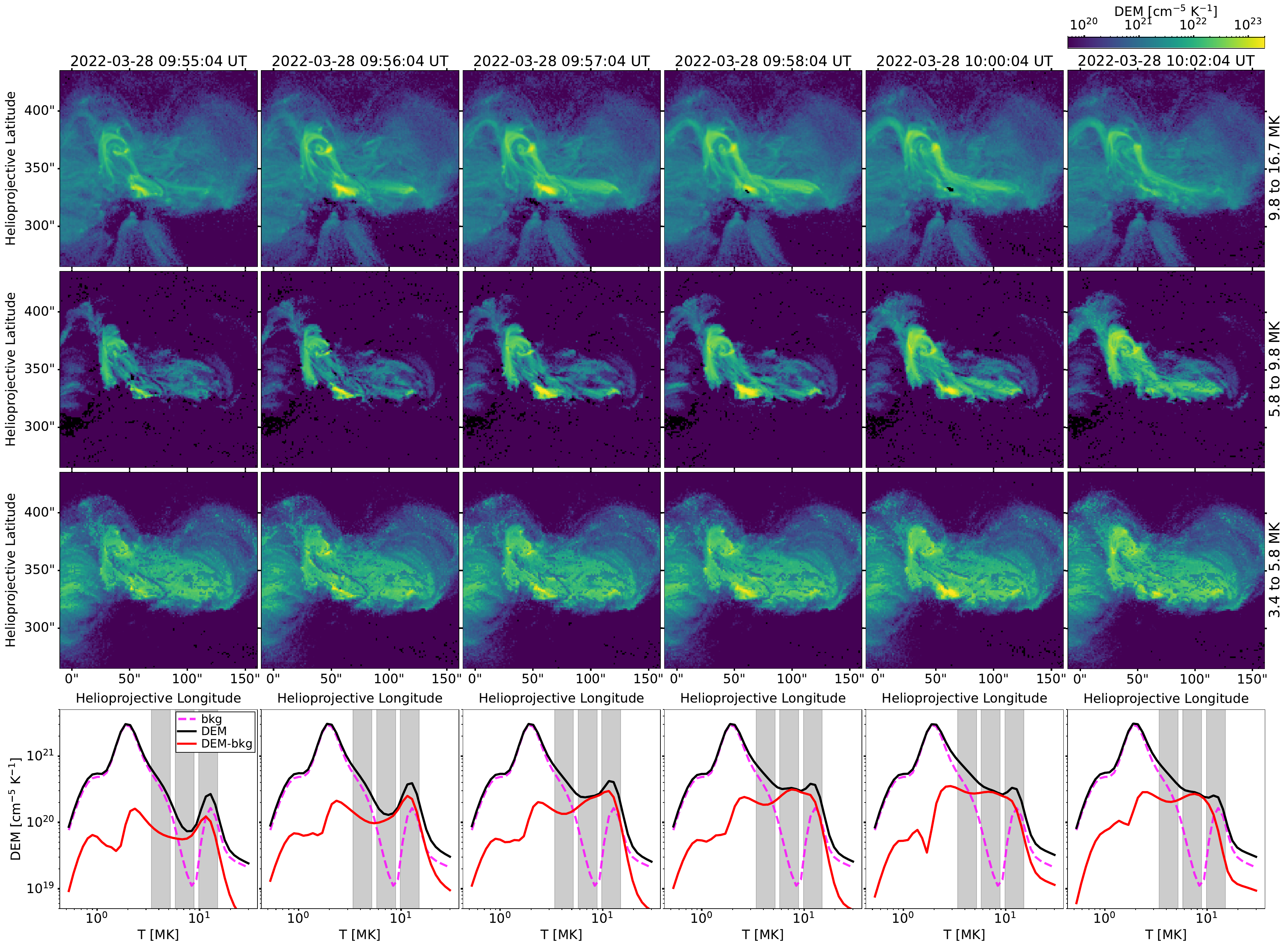}
    \caption{Overview of AIA DEM reconstructions for six selected times (columns) during the C2 flare.
    Top three rows: DEM maps showing for each pixel the mean DEM in three temperature ranges (rows): 9.8 -- 16.7 MK (first row), 5.8 -- 9.8 MK (second row), 3.4 -- 5.2 MK (third row).
    Bottom row: Mean DEM distribution (black) and background subtracted mean DEM distribution (red) of all pixels as a function of temperature. The background (bkg; pink dashed) represents the DEM distribution at 09:00~UT. The gray areas mark the temperature ranges for the DEM maps shown above.
    The associated movie is available online.}
    \label{fig:DEM_overview}
\end{figure*}

In this section, we present results from the analysis of flare-related observations. In particular, we have investigated signatures of energy release and transport through the analysis of X-ray emission observed by STIX and through DEM reconstructions of the flaring plasma derived from AIA observations. 

Figure \ref{fig:STIX_overview_multiple} gives an overview of the STIX observations and their relation to the EUI and AIA images and the 1--8~Å flux measured by the Geostationary Operational Environmental Satellites (GOES).
The preflare HXR burst visible in the STIX 4--10~keV and 10--15~keV energy bands at about 09:46~UT corresponds to the time when we observe the strongest brightening in the AIA 94 and 131~Å channels below the filament at the location of the developing flare arcade (see Fig.~\ref{fig:AIA_Cflare_overview}, second column). The counts in the lower 4--10~keV band began to increase steadily with this HXR burst, as did the GOES 1--8~Å emission, further indicating that it may have been connected to the onset of the flare. In the higher energy 15--25~keV STIX band the flare lasted only about 4 to 5 minutes, peaking at about 09:56~UT, while the 4--10~keV light curve shows a longer profile that more closely resembles the shape of the GOES observations. We divided the more impulsive phase of the flare into 40~s time intervals for further spectral and imaging analysis of its evolution and placed a sixth interval around the preflare HXR burst.

For the preflare HXR burst (09:45:35 to 09:46:15~UT) we find that the observed X-ray spectrum can be well approximated by a single isothermal model ($EM = 9.9 \times 10^{45}$ cm$^{-3}$ and $T = 12.1$ MK) up to 11~keV. At higher energies a slight excess of counts compared to the thermal model is visible but this can not be explained by a non-thermal component. For the five time intervals between 09:54:55 and 09:57:35~UT, representing the main flare, a non-thermal tick target model had to be included in addition to the isothermal model to explain the observed spectrum up to 28~keV. This fitted non-thermal spectrum remains soft throughout the flare, reaching a maximum hardness of $\delta = 8$ during the peak of the flare (09:55:35 to 09:56:15~UT). The isothermal plasma temperature modeled from the thermal emission peaks at 16.7~MK during the same 09:55:35 to 09:56:15~UT interval, while the EM increases throughout the analyzed time frame.

We note that these spectra can also be fitted well by including a second hotter isothermal component in place of the nonthermal component. For example, during the peak of the flare (09:55:35 to 09:56:15~UT) the spectrum is well approximated by a combination of a cooler 15.8~MK and a hotter 36.1~MK isothermal component. However, no such second component at temperatures above 16~MK is indicated by the AIA DEM analysis, which supports the nonthermal interpretation of the HXR power-law component (cf. Fig. 5 and discussion below).

We have constructed images for the 6-10~keV thermal part of the spectrum for all six time intervals.
Overlaying these STIX images onto the EUI/FSI images consistently places the main thermal source toward the southern part of the filament. The line of sight through the maximum of the thermal source reprojected to the AIA perspective consistently intersects the top of the small flare arcade, including during the preflare burst. Therefore, the flare loops were probably the main source of thermal X-ray emission throughout the flare. The same line of sight also intersects the hot channel, which probably contributed to the overall thermal X-ray emission, but the maximum position does not change significantly once the delayed brightening of the hot channel in EUV begins. Additional faint thermal X-ray sources, marked by the 30\% contours in the EUI images, follow an arc slightly above the filament. They may correspond to the hot loops around the northern part of the filament seen in AIA 94 and 131~Å.

For the four time intervals between 09:54:15 and 09:56:55~UT, we were also able to produce images for the higher energies covering the nonthermal part of the spectrum. We fixed the lower limit of the energy range considered for imaging at 15~keV and set the upper limit at 18, 20, 25, and 25 keV to account for the different upper extents of the spectra. For the last time interval (09:56:55 to 09:57:35~UT) we were not able to produce a satisfactory image in this higher energy part of the spectrum.
 
The reconstructed images show that the nonthermal STIX HXR emission comes mainly from a location very close to the main thermal source at the small loop arcade, but with the strongest emission consistently at a slightly higher altitude. This puts it closer to the filament breakup region and probably the region of reconnection during this flare. The emission is therefore not coming from the footpoints of the flare, whose locations are indicated by the AIA 1600~Å contours, as commonly observed in flares.

To follow the plasma heating in more detail, we produced DEM reconstructions from the AIA observations over the duration of the flare.
Figure \ref{fig:DEM_overview} shows a time series of DEM maps for three temperature ranges. For comparison, STIX is only sensitive to the highest temperature range (9.8--16.7~MK) shown in the top row.

During the onset of the flare, represented by the 09:55:04~UT DEM maps, we observe a continuous emission increase from the small loop arcade and the northern loops, mainly in the 9.8--16.7 and 5.8--9.8~MK temperature ranges. In the 9.8--16.7~MK temperature range, the small loop arcade dominates the emission. The extended channel is also visible as a thin band in the highest temperature range but its emission is absent from the lower temperature ranges.

By 09:56:04~UT, corresponding to the peak of the STIX 15--25~keV energy band, the emission has further increased in all features. The extended channel is now well visible in the highest temperature range, indicating the prevalence of heated plasma in the channel, while the small loop arcade still dominates the emission in all temperature ranges.

09:57:04~UT marks the end of the energy release according to the STIX 15-25~keV energy band. The extended channel has continued to brighten in the minute since the peak of the flare, while the emission from the small arcade loops and the northern loops remained visually at about the same level in the highest temperature range. The emission from the small loops increased in the 5.8--9.8 MK range, indicating a cooling of the flaring plasma.

Until 09:58:04~UT we observe no further emission increase in the highest temperature range. Rather, the small loops show a significant decrease in emission in the highest temperature range, while peaking in the 5.6--9.8 MK temperature range, indicating further cooling. On the contrary, the extended channel is still largely absent in the 5.6--9.8 MK temperature range, indicating that the plasma has not yet cooled significantly.

Two minutes later, at 10:00:04~UT, the small loops no longer emit in the highest temperature range, while their emission peaks in the lowest 3.4--5.8~MK range. The extended channel has also faded in the highest temperature range, but remains visible. In addition, it is now clearly visible in the intermediate temperature range, indicating that parts of this plasma have finally cooled below about 10~MK. Most of the features have faded significantly by the time the last DEM maps show 10:02:04~UT, six minutes after the peak of the flare.

The background-subtracted DEM profiles (DEM-bkg; red lines) in the bottom panels of the figure show the initial emission increase as a sharp peak above 10~MK that reaches its maximum at 09:57:04~UT. Subsequently, the emission decreases at these temperatures as the peak shifts to lower temperatures. The DEM curve drops quite steeply towards 20~MK throughout the observations.

In summary, the DEM results appear to be consistent with the STIX observations, which associate the strongest thermal emission with the small loop arcade. The heating of the hot channel lags behind the loop arcade, but is consistent with the time of energy release suggested by the STIX 15--25~kev light curve and with the AIA observations, which show that the breakup of the original filament continued until about this time. Looking also at the evolution of the DEM distributions, we find no evidence for a significant hot component above 30~MK, as would be suggested by a second isothermal fit to the STIX spectrum. The errors associated with the used DEM reconstruction method and the results presented in this section are discussed in the appendix.

\subsection{NLFF extrapolations of the pre-flare magnetic field}\label{sec:results_NLFF}

\begin{figure*}
\centering
  \includegraphics[width=18cm]{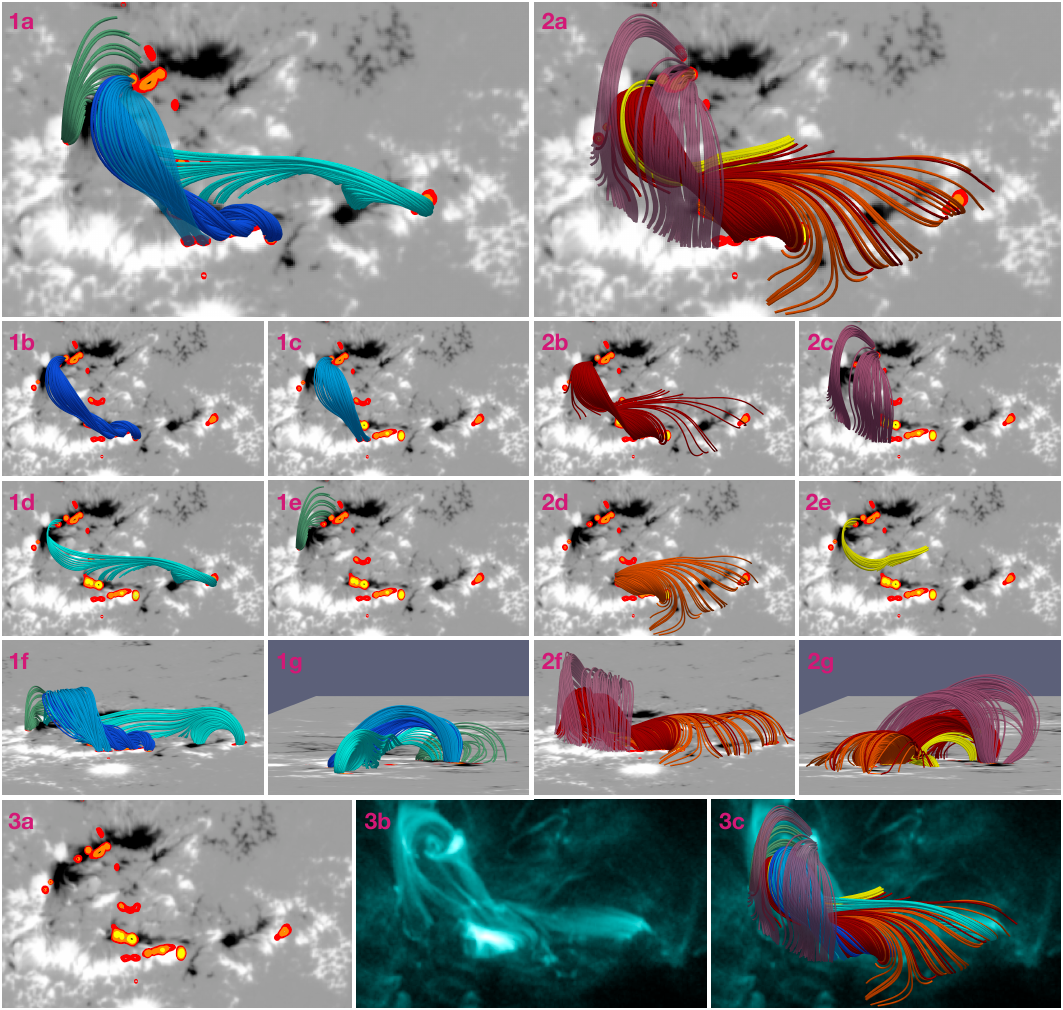}
    \caption{Field lines from a pre-flare magnetic field reconstruction (09:36~UT) associated with the AIA 1600~Å enhancements during the C2 flare. The background consists of the HMI magnetogram in CEA projection from 09:36~UT with contours from an AIA 1600~Å image taken at 09:55:47~UT corresponding to 50, 100, and 200 counts/s.
    Panel 1a: All field lines with seed points within a positive polarity region. 
    Panels 1b to 1e: Individual display of selected field lines. 
    Panels 1f and 1g: Combined field lines shown from two different perspectives.
    Panels 2a to 2g: The same structure but for field lines with seed points in negative polarity regions.
    Panel 3a: The background HMI image with AIA 1600 Å contours without any field lines.
    Panel 3b: An AIA 131~Å image taken at 09:56~UT during the peak of the flare.
    Panel 3c: The combined structure of all field lines displayed above the AIA image from panel 3a.}
    \label{fig:NLFF_overview}
\end{figure*}

\begin{figure}
  \resizebox{\hsize}{!}{\includegraphics{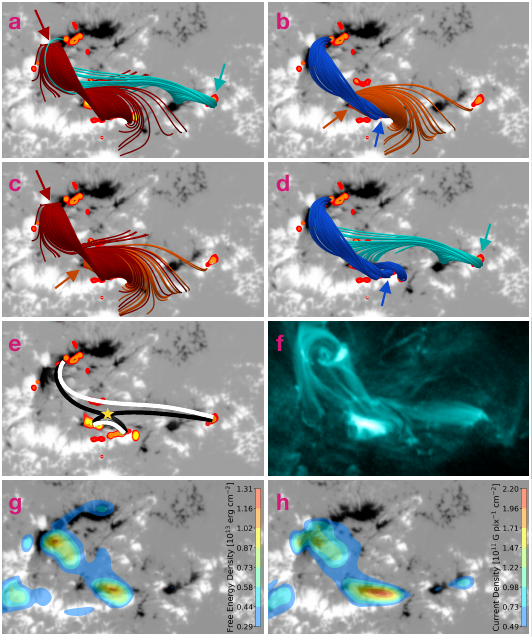}}
  \caption{Overview of the proposed reconnection scenario.
  Panels a to d: Pairs of field structures starting from four relevant regions: The flare ribbons in the major positive and minor negative polarity regions and the kernels in the major negative and western positive polarity regions. The arrows indicate the region of the seed source for starting the field lines of the corresponding color.
  Panel e: Illustration of the proposed reconnection scenario. Black field lines represent field lines from the pre-flare configuration that were reconnected to form the white field lines during the flare. The reconnection region is indicated by the yellow star.
  Panel f: AIA 131~Å image taken during the peak of the flare at 09:56 UT.
  Panel g: Filled contours of the vertically integrated free energy density during the flare (09:58~UT).
  Panel h: Filled contours of the vertically integrated current density during the flare (09:58~UT).}
  \label{fig:NLFF_reconnection}
\end{figure}

\begin{figure}
  \resizebox{\hsize}{!}{\includegraphics{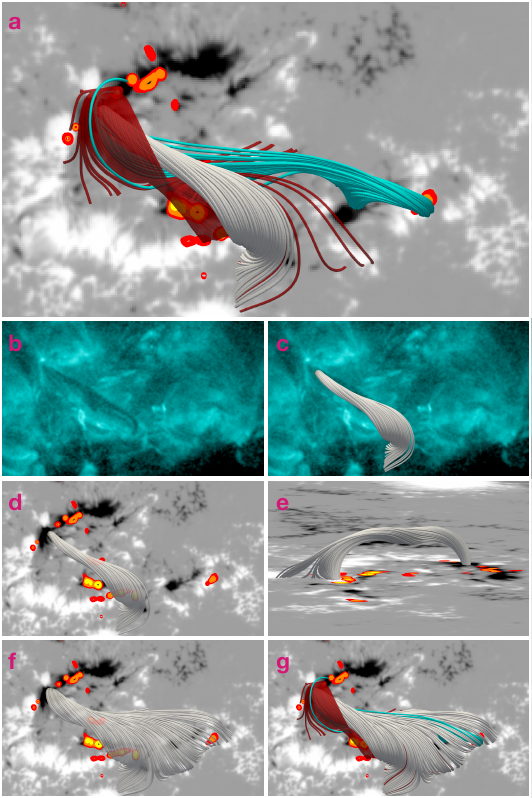}}
  \caption{The filament channel and its relation to other field structures.
  Panel a: Field lines corresponding to the original filament channel (gray) together with field structures involved in the proposed reconnection process (see Fig.~\ref{fig:NLFF_reconnection}). The filament field lines were started in the north. 
  Panel b: AIA 131~Å image taken at 09:36~UT showing the original filament. 
  Panel c: Filament field structure on top of the AIA image from panel b. 
  Panels d and e: The filament field structure on its own from a top-down and side-on perspective. 
  Panel f: The extended filament structure revealed by a larger seed source at the same location.
  Panel g: Relation of the extended filament structure to the same field lines as in panel a.}
  \label{fig:NLFF_filament}
\end{figure}

In this section, we present our analysis of a NLFF extrapolation of the magnetic field before the restructuring of the filament (09:36~UT) using the method of \citet{Jarolim2023}. We evaluate the quality of the extrapolation with standard metrics for quantifying force-free solutions (see Sect. \ref{sec:methods_NLFF}). We find that the divergence is sufficiently low ($L_{div,n}(\vec{B}) = 0.004 \textrm{ pixel}^{-1}$) and that the separation between magnetic field vector  $\vec{B}$ and the current density $\vec{J}$ is a good approximation of the force-free condition ($\sigma_J = 0.19$, $\theta_J = 10.8^\circ$). Despite the extended field-of-view of SHARP 8088, the metrics are in a similar range as previous applications of physics-informed neural networks for magnetic field extrapolations \citep{Jarolim2023,Jarolim2024}.

Figure \ref{fig:NLFF_overview} gives an overview of selected pre-flare magnetic field structures that start within the contours of the AIA 1600~Å brightenings during the C2 flare, which is indicative that these field structures became likely involved in the reconnection at that time. We divide them into two groups according to the magnetic polarity in which their seeds are located. Field structures originating from AIA 1600~Å brightenings in a positive polarity region are shown collectively in panel 1a and separately in panels 1b to 1e. In panels 1f and 1g we show the combined field structures from two additional vantage points located south and west of the AR, respectively. Panels 2a through 2g follow the same structure, but focus on field structures starting from AIA 1600~Å brightenings that lie within negative polarity regions.

Panel 1b shows field lines extending from the J-shaped AIA 1600~Å flare ribbon in the major positive polarity region. Field lines to the west of this region appear to connect to the nearby minor negative polarity region in the form of small loops, while the eastern parts of this ribbon are connected to the major negative polarity region. 
Panel 1c shows that the field lines starting at the eastern end of the J-shaped ribbon merge into an overlying arcade that lies above the dark blue field lines of panel 1b. 
Panel 1d shows the field structures seeded by the AIA 1600~Å kernel in the western positive polarity region, corresponding to the footpoint of the hot channel during the flare. This region was connected to an adjacent negative polarity, the minor negative polarity region, and the major negative polarity region. The shape of the longest field lines, extending toward the major negative polarity region, already followed the shape of the new filament. This indicates that part of the new filament structure already existed before the C2 flare, but did not contain cool plasma.
Panel 1e shows another set of field lines in the north of the AR that were probably not important for understanding the filament restructuring, but may have played a role in the formation of the bright loops surrounding the northern part of the filament during the C2 flare.

Panel 2b shows a field structure originating from an AIA 1600~Å kernel in the major negative polarity region, where the field lines transition from an overlying arcade to a strong S-shape. The anchor points of the field lines follow a J-shape in the major positive polarity region, which does not exactly follow the J-shaped ribbon in AIA 1600~Å as one might expect. This may be due to the limitations of the NLFF magnetic field reconstruction. In addition, the structure contains even more sheared field lines that connect to the positive polarity region in the west. This again shows that magnetic field lines following the shape of the new filament were already present in the pre-flare magnetic field. It also shows that the more sheared field lines lie below the less sheared ones, suggesting that the new filament channel lay below the original channel.
 
The remaining AIA 1600~Å kernels in the major negative polarity region are associated with overlying loop structures, shown in panel 2c. They may have been relevant for the bright loop structures in the north. Some of their southern footpoints nicely match brightness enhancements in AIA 131~Å, as can be seen in panels 3b and 3c.
Panel 2d shows structures starting from the 1600~Å flare ribbon in the minor negative polarity region. They connect to parts of the J-shaped ribbon in the major positive polarity region and then transition to the western positive polarity region surrounding the new filament footpoint location.
The field lines shown in panel 2e start from or cross over an AIA 1600~Å enhancement in a very weak magnetic region. They start as part of the general filament structure in the north but do not connect to the minor negative polarity. They were also revealed as part of the red field structure in panel 2b.

Panels 3b and 3c show an AIA 131~Å image taken at the peak of the flare (09:56~UT) and the same image overlaid with the combined field structures of both polarities. The general structure of the pre-flare NLFF magnetic field follows the shape of all the major brightenings in AIA 131~Å. This is a good indication that our method of starting field lines from the contours of the AIA 1600~Å brightenings reveals field structures that were reconnected or otherwise activated during the C2 flare.

Figure \ref{fig:NLFF_reconnection} focuses on those field structures that were likely involved in the magnetic reconnection that resulted in the small flare loop arcade and the extended hot channel. The four relevant AIA 1600~Å brightenings are those of the flare ribbons in the major positive and minor negative polarity regions and the kernels in the major negative and the western positive polarity region (Fig.~\ref{fig:NLFF_overview}, panels 1b, 2b, 2d, and 1d, respectively). We show four pairs of field structures (panels c through f) obtained by matching any two of these relevant footpoints from opposite polarity regions. All pairs show that there are generally two relevant field systems. One connects the major positive and negative magnetic polarity regions with S-shaped field lines surrounding the original filament channel, whose footpoints follow a J-shape in the positive polarity region (dark red and blue field lines). The other relevant field system connects the minor negative polarity region with the western positive polarity region around the new filament footpoint (turquoise field lines). The reconnection between the field lines of the two systems could explain both the small flare loops and the extended hot channel. An illustration of this process is shown in panel e.

Panel g shows that there were two regions of high free magnetic energy density at the time of the flare. The northern one corresponds to the northern filament footpoint (see also Fig.~\ref{fig:NLFF_filament}), while the southern one corresponds to the proposed reconnection region above the minor PIL between the major positive and minor negative polarity regions. Comparing this with panel h, we see that the strongest currents at the time of the flare are also located around this southern region. This indicates the formation of a current layer above the PIL due to the shearing and converging motions around it, which are discussed in the subsequent subsection. Such a current layer is required for the onset of magnetic reconnection between the adjacent flux systems, and its steepening supports the onset of reconnection \cite[e.g.,][]{Mikic1994}. Moreover, this reconnection region would be consistent with the position derived for the high-energy X-ray emission observed by STIX (cf., Fig.~\ref{fig:STIX_overview_multiple}).

In Fig.~\ref{fig:NLFF_filament} we show gray field lines better representing the actual filament channel. They were revealed by placing a seed source on the inside of the northern bend of the dark red S-shaped field lines. In panel a, we see how the original filament extended as a twisted channel from this starting point to the inside of the J-shaped ribbon in the south. Comparison with a pre-flare AIA 131~Å image taken at 09:36~UT in panels b and c confirms that these field lines follow the path of the original filament as observed in EUV. In panels d and e one can get a better sense of the relationship between the filament structure, the HMI LOS magnetic field, and the AIA 1600~Å kernels. In addition, panel e provides a side-on view that shows a clear dip in the twisted channel that could serve for the suspension of cool plasma in the corona. However, the EUI observations in Fig.~\ref{fig:EUI_Cflare_overview} suggest an arched shape of the filament without dips. Increasing the radius of the seed source (panels f and g) reveals more of the field structure surrounding the original filament. The field lines transition smoothly into the longer and more sheared new filament channel that wraps around and then crosses under the original filament. The field structure suggests that the topology of the reconnection sketched in Fig.~\ref{fig:NLFF_reconnection} panel e can also explain the restructuring of the filament during the flare. When the process proceeds from the dark red and turquoise field lines to reconnection between the inner twisted and surrounding more sheared parts of the gray field lines, the filament would break up and reform as observed.

\subsection{AR emergence and original filament formation}\label{sec:results_emergence}

\begin{figure*}
\centering
  \includegraphics[width=18cm]{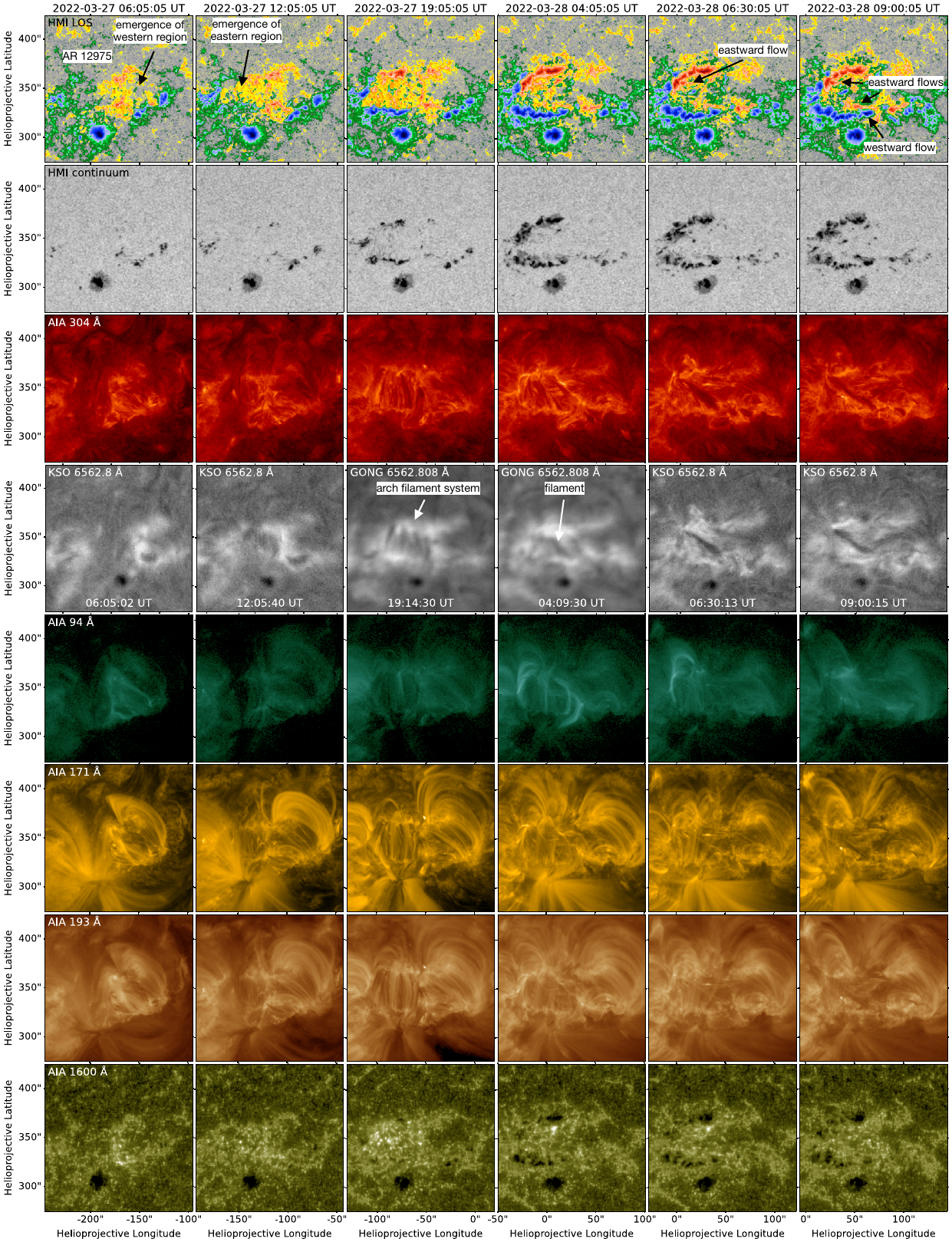}
    \caption{Emergence of AR 12975 and the formation of the original filament as observed by SDO/HMI, selected SDO/AIA channels, and by KSO and GONG in H$\alpha$ at six times (columns). HMI LOS magnetograms are scaled from $-1500$~G (red) to $+1500$~G (blue). The associated movie is available online.}
    \label{fig:AIA_emergence_overview}
\end{figure*}

\begin{figure*}
\centering
  \includegraphics[width=18cm]{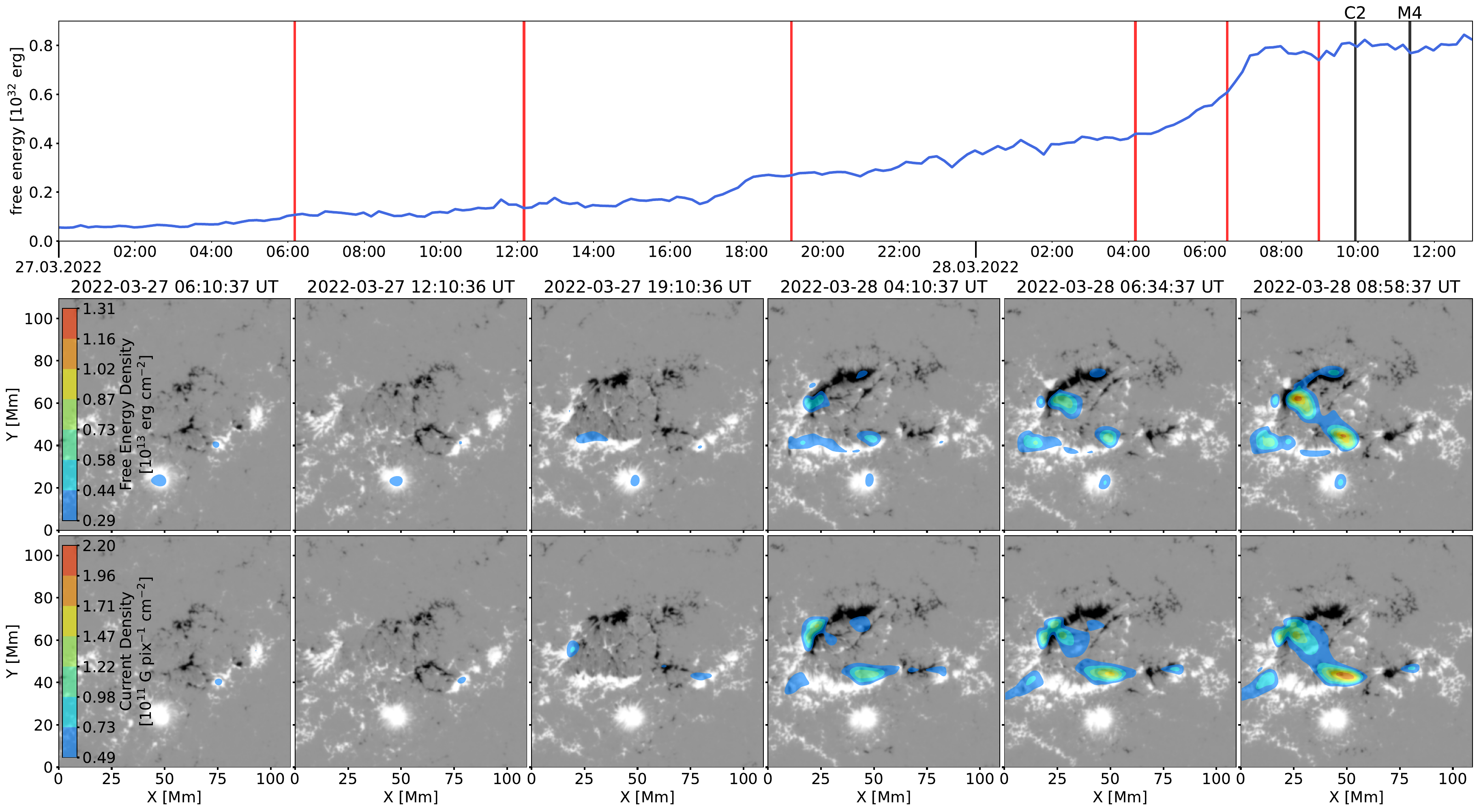}
    \caption{Evolution of free magnetic energy and current density during the formation of AR 12975 and the original filament. The top panel shows the time evolution of the total free magnetic energy in the submaps shown below, with their observation times marked with red lines. The peak times of the C2 and M4 flares are marked with black lines. The submaps in the bottom two rows show the HMI magnetogram in CEA projection scaled to $\pm1000$~G and overplotted with filled contours of either the free magnetic energy density (middle row) or the current density (bottom row) integrated along the z-axis. The associated movie is available online.}
    \label{fig:free_energy_and_current_overview}
\end{figure*}

This section gives an overview of the evolution of AR 12975 from the first flux emergence to the subsequent formation of the original short filament. Figure \ref{fig:AIA_emergence_overview} shows the AR over about one day as observed in (E)UV by AIA, in H$\alpha$ by either KSO or a GONG telescope if the former was not available, and in the continuum by HMI. In addition, the evolution of the LOS magnetic field as measured by HMI is shown in the upper panels. During this period the AR evolved rapidly, and the processes discussed below can be followed much better in the movie accompanying the figure, which also gives a more complete overview of the full evolution. Figure \ref{fig:free_energy_and_current_overview} gives a complementary overview of the evolution of the vertically integrated free magnetic energy and current densities as derived from our NLFF extrapolations during AR emergence and filament formation, focusing on the same times for easy comparison. The exact observation times may differ by up to $\pm$6 minutes due to the 720~s cadence of the HMI vector magnetograms used for the NLFF extrapolations.

The first flux emerges in the west of the AR throughout 26 March 2022 (see movie). The first observations included in Fig.~\ref{fig:AIA_emergence_overview} are from 06:05:05~UT on 27 March 2022 and show the bipolar configuration with connecting loops between the two polarities. By 12:05:05~UT, the formation of the eastern part of the AR has begun, which is characterized by the emerging flux flowing away from the PIL and accumulating as two elongated polarity regions in the north and south.

At 19:05:05~UT on 27 March 2022, the main outline of the AR discussed in previous sections is already recognizable. One can clearly distinguish the major positive and negative polarity regions, which are connected by an ordered loop arcade in EUV and form an arch filament system in H$\alpha$ aligned perpendicular to the PIL. During this period of strong flux emergence, the region between the two major polarities shows an overall weak negative polarity with numerous AIA 1600~Å bright kernels around the remaining spots of positive polarity indicating flux cancelation.

By 04:05:05~UT on 28 March 2022, more flux has emerged, strengthening the two main polarity regions. The movie shows how the loop arcade develops more sheared sections up to this point. This period is also characterized by a high level of activity in channels sensitive to hot plasma (represented by AIA 94~Å in Fig.~\ref{fig:AIA_emergence_overview}), with different parts of the loop system constantly brightening, possibly indicating reconnection within the arcade. The images included in the figure represent the first time the filament becomes clearly visible. In the previously weakly negative polarity region between the major polarities, some positive flux has begun to appear. 

Below the major negative polarity in the north, we observe patches of negative flux flowing eastward parallel to the PIL. In addition, we observe an apparent clockwise rotation of this flux as it becomes part of the major negative polarity region. This motion can also be seen as photospheric flows in the HMI continuum series. The location where the negative flux settles is consistent with the location of the northern filament footpoint, and the onset of this flow coincides well with the emergence of the filament from the arcade. Comparison with Fig.~\ref{fig:free_energy_and_current_overview} shows that this time also marks the beginning of the rapid increase in free magnetic energy leading up to the C2 and later M4 flares, and that a source of concentrated free magnetic energy density begins to form at the location of the northern filament footpoint and the final location of the streaming flux. A second region of free magnetic energy develops between the major positive polarity and the adjacent minor negative polarity region, which also begins to form at this time.

At 06:30:05~UT (Fig.~\ref{fig:AIA_emergence_overview}) the arcade loop structures surrounding the filament have begun to fade, making the filament more visible. The motion of the negative flux patches parallel to the PIL and some rotation continues as the northern filament footpoint appears to be moving further northeast in the AIA EUV observations. The free magnetic energy at the northern filament footpoint and between the major positive and minor negative polarity regions now dominates the free magnetic energy stored in the AR and continues to increase rapidly (Fig.~\ref{fig:free_energy_and_current_overview}).

By 09:00:05~UT the filament has further increased its shear, mainly due to a strong westward motion of the southern footpoint driven by the westward drift of the western tip of the positive polarity region, where the filament is rooted. At the same time, the adjacent minor negative polarity region continues to develop and flow eastward. Such antiparallel flows build up and steepen (narrow) an essentially vertical current layer between the increasingly sheared coronal flux systems rooted in these photospheric polarities \cite[e.g.,][]{Mikic1994}. Additionally, the ongoing emergence of flux continues to push the minor negative polarity southward, toward the adjacent major positive polarity. This represents a flow converging at the PIL between them and further contributes to the steepening of the current layer and the associated buildup of free magnetic energy (Fig.~\ref{fig:free_energy_and_current_overview}), which ultimately causes the reconnection. This is the starting point for the filament evolution discussed in detail in this paper and corresponds to the time shown in the first column of Fig.~\ref{fig:AIA_Cflare_overview}.

\section{Discussion}\label{sec:discussion}

We have studied in detail the rapid evolution and restructuring of the filament in AR 12975 associated by a confined C2 flare at 09:56 UT on 28 March 2022, taking advantage of the favorable configuration between STIX and EUI onboard Solar Orbiter and AIA and HMI onboard SDO. Solar Orbiter was near its first science perihelion, at a position 0.33 AU from the Sun and $83.5^\circ$ west of the Sun-Earth line. From this perspective, the filament appeared near the eastern limb, giving STIX and EUI a close-up side-on view of the flare signatures in HXR and of the filament's height evolution. AIA and HMI offered a complementary on-disk view that captured the restructuring of the filament in greater detail and provided ideal conditions for the NLFF magnetic field extrapolations. These observations were further supplemented by H$\alpha$ observations from KSO and GONG.

During the C2 flare, the southern half of the filament disappeared and plasma subsequently flowed into a new, longer filament channel with a different positive footpoint and a topology very similar to an EUV hot channel observed during the flare. In addition, we presented an overview of the emergence and evolution of the AR and the quick formation of the original filament from an arch filament system over about one day. Combined with the study of \citet{Purkhart2023}, focusing on the M4 flare associated with the eventual filament eruption at about 11:20~UT on 28 March 2022, about 1.5 hours after the C2 flare, this provides a complete overview of the filament's evolution from emergence to eruption.

Our analysis revealed the rapid evolution of the AR and the filament (see Sect.~\ref{sec:results_emergence}) on the day prior to the studied event. The western part of the AR emerged first, followed by the emergence of the two major polarities in the east, initially connected by an arch filament system. Over about half a day, the shear in parts of this arcade gradually increased, while at the same time, most of the arch filament system disappeared, revealing the original filament. The increase of shear was mainly driven by eastward flows, parallel to the main PIL of the emerged flux, of the main negative polarity, where the filament was rooted. The flux in the footpoint area also showed a clockwise rotation. In addition, we observed numerous brightenings of the arcade loops in AIA 94 and 131~Å and the emergence of positive flux below the filament, accompanied by numerous AIA 1600~Å kernels. These observations suggest a high rate of (small-scale) reconnection events within the arcade, which presumably have contributed to the formation of the original filament channel from the emerged arcade field. In addition, the observed footpoint rotation builds up twist in the coronal flux holding the filament.

After its emergence, the filament continued to increase in shear due to antiparallel motions of the filament footpoints, driven by the ongoing photospheric flows. During the last hours before the restructuring, the particularly strong westward drift of the southern filament footpoint in the major positive polarity was accompanied by the formation and eastward drift of the adjacent minor negative polarity region, resulting in antiparallel flows along this secondary but high-gradient PIL. The NLFF extrapolations reveal a corresponding strong buildup of free magnetic energy at the northern filament footpoint and around the PIL near the southern filament footpoint during this period.

Our analysis of the pre-flare NLFF magnetic field at 09:36~UT on 28 March 2022 in Sect.~\ref{sec:results_NLFF} showed that field lines following the initial and the restructured filament channels probably existed as part of an extended structure prior to the C2 flare. We identified the original filament channel in our NLFF reconstructions as twisted field lines extending between the eastern end of the major negative polarity region and the western end of the major positive polarity region (see Fig.~\ref{fig:NLFF_filament}, panels a to e). The northern footpoint corresponds to the location where the photospheric flows parallel to the PIL settled into the major negative polarity region and where we observed the rotation, while the southern footpoint lies very near (slightly south of) the observed footpoint in the major positive polarity region. The field lines associated with the new filament channel were twisted around the original channel in the north, extended below it in the middle part where the reconfiguration occurred, and continued to a minor positive polarity west of the main positive polarity (see \ref{fig:NLFF_filament} panels f and g).

Considering the field structures originating from the location of the AIA 1600~Å flare ribbons and kernels, we propose that the reconnection leading to the EUV hot channel and the small flare loop arcade and its associated ribbons probably occurred between field lines wrapping around the original filament channel and field lines connecting the minor negative polarity to the western positive polarity region (see Fig.~\ref{fig:NLFF_reconnection}). Such a reconnection scenario could explain the observations of both the small flare arcade and the bright elongated channel (panel e). The proposed reconnection location above the PIL between the major positive and minor negative polarity regions was one of two regions storing most of the free magnetic energy at that time (panel g), is consistent with the location of the strongest currents at the time of the flare (panel h), and agrees with the inferred source location of the non-thermal X-ray emission observed by STIX (see Fig.~\ref{fig:STIX_overview_multiple}). In addition, the antiparallel and converging flows observed along this PIL provide a mechanism for pushing the involved field systems together, building up an essentially vertical current layer and its associated free energy that eventually enables the fast reconnection during the C2 flare.

The reconnection scenario suggested in Fig.~\ref{fig:NLFF_reconnection} panel e is the loop-loop reconnection scenario for flares originally proposed by \citet{Nishio1997} and \citet{Hanaoka1997}. This is a model for confined flares that do not involve any significant rise of the involved flux, like in the event considered here. The model predicts that the loops exchange their footpoints to form two new loops. This differs from the classical model of ``flare'' reconnection in an essentially vertical current sheet, which forms an arcade of flare loops as the lower reconnection product and adds flux to a flux rope with different footpoints as the upper reconnection product. However, the observations and our NLFF magnetic field reconstruction show that the lower product of loop-loop reconnection can actually be similar to the one of classical flare reconnection. The reason is that the original ``loops'' can have a sheet-like geometry around their contact point. In the present event, the reconnecting flux systems are rooted in essentially linear and closely spaced photospheric flux concentrations: the main positive polarity in the south and the minor negative polarity adjacent to the north. The contact region is a horizontally extended current layer (Fig.~\ref{fig:NLFF_reconnection}, panel h), leading to a reconnection geometry as in the standard flare model and to a flare loop arcade with its associated pair of flare ribbons. At the same time, the loop-loop reconnection during the C2 flare can also be regarded as a tether-cutting reconnection, because one of the resulting loops is much longer than each of the original ones (while the short resulting loops in the small arcade stay passive in the further evolution of the active region). This is discussed further below.

Confined flares due to loop-loop reconnection have been termed ``Type~I'' confined flares by \citet{LiT2019}. These authors also found that typically arcades of flare loops are formed and that these arcades typically possess high shear, different from ``Type~II'' confined flares, which are regular flux eruptions subsequently stopped in the corona by the overlying flux. The fact that typically arcades of flare loops are seen in place of the small resulting loop in the original model is probably a simple result of the higher resolution of the AIA data. The event considered here shows that the shear of the loop arcade in Type~I confined flares can be rather moderate, albeit not very small. This difference from the finding in \citet{LiT2019} simply results from the geometry of the reconnecting flux systems, whose footpoints adjacent to the reconnecting current sheet (AIA 1600~Å ribbons) are only moderately displaced along the PIL (Figs. \ref{fig:NLFF_overview} and \ref{fig:NLFF_reconnection}). 

It seems likely that the original filament channel was even more directly involved in the reconnection than suggested by our NLFF extrapolation. In the AIA observations, the southern filament footpoint seems to be anchored directly or very close to the brightest AIA 1600 Å emission kernel at the hook of the J-shaped ribbon (see Fig.~\ref{fig:AIA_Cflare_overview}, 09:56:18~UT, top panel). However, the J-shaped bend formed by the footpoints of the field lines in our NLFF extrapolation does not coincide with the J-shaped flare ribbon but is located further east (see Fig.~\ref{fig:NLFF_overview}, panel 2b). Instead, the brightest part of the ribbon corresponds to closed field lines in our reconstructions (see Fig.~\ref{fig:NLFF_overview}, panel 1b). It seems likely that this is due to a shortcoming of the NLFF model since we are modeling a high pressure and density filament which is neglected in the NLFF setting. The actual bend in the field lines and the filament footpoint within were probably situated further to the west, in agreement with the AIA 1600 Å and EUV observations.

In addition, AIA observations show the rapid and complete disappearance of the filament section south of the reconnection region during the C2 flare. In this phase, the edges of the remaining plasma facing the reconnection region were heated, clearly visible as the bright edge in the north, but also in the disappearing southern plasma (see Fig.~\ref{fig:AIA_Cflare_overview} and accompanying movie). The brightening in the middle of the filament and the disappearance of the southern part was a process that progressed from the underside through the whole cross-section of the original filament. Both are consistent with loop-loop reconnection similar to the one indicated by the EUV hot channel, but now involving the filament channel itself and nearby ambient flux rooted in the western positive polarity region, i.e., passing under the middle of the original filament (gray field lines in Fig.~\ref{fig:NLFF_filament}). Such reconnection could also make plausible the unusual absence of HXR footpoint sources because particles accelerated in the reconnection region might already be stopped by the dense filament plasma near the reconnection site. Indeed, STIX may have observed soft ($\delta = 8$) non-thermal emission directly from the reconnection region (see Fig.~\ref{fig:STIX_overview_multiple}). Our DEM investigation (Sect.~\ref{sec:results_flare}) concluded that an alternative explanation of the STIX spectrum with a second, super-hot 35~MK component is unlikely.

In the scenario described, the cool, dense plasma in the northern part of the original filament could flow directly into the new channel after reconnection, while the original connection to the major positive polarity would have been lost. This scenario seems to be the most consistent with the observations. In the AIA and EUI observations we see that the existing filament plasma from the north, including the heated plasma very close to the reconnection region, flowed into the new channel immediately after the flare. The drop in filament height around the breakup region (Figs. \ref{fig:EUI_Cflare_overview} and \ref{fig:EUI_Jmap}) is also consistent with the scenario that more sheared field lines under the original filament formed the new filament channel into which the plasma from the northern part subsequently flowed. The new channel was filled in the course of about 20 min, during which the flowing filament plasma showed no signs of an intact connection to the original footpoint. This is also true during the later M4 flare, when the filament plasma drains towards the new footpoint, driven by the upward whipping northern footpoint. However, the small loop arcade continued to grow during the filament eruption, suggesting that field lines with similar topology to those reconnected during the C2 flare were still present at this later time. This was also the result of the NLFF extrapolation performed by \citet{Purkhart2023} for the pre-eruptive state of the M4 flare at 11:00~UT.

The general consequence of the proposed loop-loop reconnection is that the new filament channel grows while the original channel is simultaneously decaying. This also represents a sort of tether-cutting scenario by creating a long filament channel from two shorter flux systems. This decreases the line tying of the new filament channel compared to the original one. An interesting question is why this did not lead to an immediate runaway reconnection driving an eruption during the C2 flare.

It appears that the tether-cutting process has played an important role throughout the evolution of the region. Brightenings at the location of the small loop arcade were already observed well before the C2 flare in AIA, with the brightest one (see Fig.~\ref{fig:AIA_Cflare_overview}, first column) commencing the onset of the flare and associated with a HXR peak (Fig.~\ref{fig:STIX_overview_multiple}, first column). Further brightenings were observed throughout the slow rise and destabilization of the filament leading up to the M4 flare. The C2 flare marked a period of fast reconnection in this ongoing process, during which large parts or all of the original filament were reconnected. One of the most important consequences of this ongoing process was the formation of the long filament channel through reconnection involving field lines passing under the filament where the breakup occurred. Thus, tether-cutting probably played a major role in pushing the filament further toward its eventual instability. However, this does not mean that the eventual eruption was driven by the tether-cutting reconnection. The process did not develop a runaway characteristic during the C2 flare (Fig.~\ref{fig:EUI_Jmap}), and only after the filament structure continued to rise did the filament finally reach its instability. This may have been the onset of another instability, like the torus-instability \citep{Kliem2006}.

Additional upward magnetic pressure could come from the twist in the filament channel \citep{Torok2003}. However, observations and NLFF extrapolations suggest that the filament under study was a weakly twisted flux rope, with about one field line turn in our NLFF extrapolation. Some of this twist was probably built up during the filament formation, as indicated by the northern footpoint rotation. The evolution of the strands in the north (see Fig.~\ref{fig:EUI_Cflare_overview}) also shows the twisted structure. Finally, at the beginning of the eruption, the twist in the long filament structure is clearly visible (e.g. 11:10~UT in the movie accompanying Fig.~\ref{fig:AIA_Cflare_overview}).

\section{Conclusion}\label{sec:conclusion} 

Our multi-point study focused on the rapid filament restructuring in AR~12975 during a confined C2 flare on 28 March 2022 at 09:56 UT, during which we observed a breakup of the original filament channel which led to the separation from its southern footpoint and the flow of plasma from the northern part into a new, longer channel with a geometry similar to an EUV hot channel observed during the flare. We combined observations from two different vantage points, taking advantage of the quasi-quadrature position of SDO and Solar Orbiter during this time.

We showed that the magnetic structures for both filament channels probably already existed in the pre-flare magnetic field, with the new longer channel extending below the middle part of the original shorter channel and then wrapping around and passing over its northern part. Based on the field structures associated with the AIA 1600~Å ribbons and kernels during the flare, we propose that the magnetic reconnection involved field lines surrounding and passing below the short filament channel and field lines closely following the southern part of the longer channel. Both field systems intersected above the minor PIL between the major positive and the adjacent minor negative polarity region. The topology represents a loop-loop reconnection scenario that realizes the concept of tether cutting and can explain both the hot channel and the small flare loop arcade observed during the flare and led to the breakup of the original filament as the reconnection progressed. This location of the reconnection region is further supported by a strong concentration of currents and free magnetic energy in this region, by the high-energy (non-thermal) HXR emission observed by STIX above the flare arcade, and by the presence of antiparallel photospheric flows along this PIL.

Observations of ongoing brightenings and even HXR emission at the site of the small flare arcade before the C2 flare, and also between the C2 flare and the filament eruption accompanied by an M4 flare about 1.5 hours later, suggest that the tether-cutting process was active throughout the whole filament rise phase. The C2 flare marked a period of fast reconnection during the otherwise more steady process, during which most of the original filament channel was probably reconnected and joined the long channel. This allowed the remaining filament plasma in the north to flow directly along the new channel after the flare. A direct consequence of this reconnection process is the buildup of the long filament channel and we conclude that tether-cutting was likely an important driver in bringing the filament closer to its eventual eruption.

The investigated event illustrates that the classical loop-loop reconnection scenario for flares \citep{Nishio1997, Hanaoka1997} can be part of a long-lasting tether-cutting reconnection process. It also shows that such reconnection can form a flare loop arcade if the contact region of the interacting loop systems has a sheet-like geometry similar to a flare current sheet and that even confined precursor flares of Type~I can contribute to the evolution toward a full eruption producing a CME.

In addition to our analysis of the filament restructuring and C2 flare, we presented an overview of the rapid emergence and evolution of the AR during the day before. The emergence of the original filament from a sheared arcade was accompanied by emerging flux, flows parallel to the PIL, footpoint rotations, and signs of reconnection in the arcade. 

Our study reveals insights into the rapid evolution of the filament driven by reconnection with a nearby field structure and discusses the consequences for its instability and eruption. Together with the analysis of the M4 flare that accompanied the eventual filament eruption by \citet{Purkhart2023}, this study provides a complete overview of the filament's lifetime and highlights the importance of multi-point, multi-instrument observations in unraveling complex solar phenomena.

\begin{acknowledgements}
Solar Orbiter is a space mission of international collaboration between ESA and NASA, operated by ESA.
The STIX instrument is an international collaboration between Switzerland, Poland, France, Czech Republic, Germany, Austria, Ireland, and Italy. 
The EUI instrument was built by CSL, IAS, MPS, MSSL/UCL, PMOD/WRC, ROB, LCF/IO with funding from the Belgian Federal Science Policy Office (BELSPO/PRODEX PEA 4000134088); the Centre National d’Etudes Spatiales (CNES); the UK Space Agency (UKSA); the Bundesministerium für Wirtschaft und Energie (BMWi) through the Deutsches Zentrum für Luft- und Raumfahrt (DLR); and the Swiss Space Office (SSO).
H$\alpha$ data were provided by the Kanzelhöhe Observatory, University of Graz, Austria.
Data were acquired by GONG instruments operated by NISP/NSO/AURA/NSF with contribution from NOAA.
This research was funded in part by the Austrian Science Fund (FWF) 10.55776/I4555. For the purpose of open access, the author has applied a CC BY public copyright license to any Author Accepted Manuscript version arising from this submission.
S.K. acknowledges the Swiss National Science Foundation Grant 200021L\_189180 for STIX.
B.K. acknowledges support by the DFG and by NASA through Grants No. 80NSSC19K0082 and 80NSSC20K1274.
\end{acknowledgements}

\bibliographystyle{aa} 
\bibliography{bib-file}

\begin{appendix}
\label{sec:appendix}

\section{DEM error analysis}


The errors associated with the regularized inversion algorithm we used for our DEM analysis are discussed in detail in the paper by \citet{Hannah2012}, which first introduced the method, and in the paper by \citet{Hannah2013}, which used an updated, faster version of the code specifically designed to efficiently handle the large number of separate DEM calculations required for high-resolution AIA images.
An advantage of this method is that it provides two errors for the reconstructed DEM in each temperature bin. A vertical error for the DEM value and a horizontal error representing the temperature resolution \citep{Hannah2012}.

To compute the vertical errors, the original code used a Monte Carlo approach \citep{Hannah2012}, while the optimized version we used for this study simply takes the linear propagation of the errors associated with the AIA source data (count statistics, background, and instrumental uncertainties) \citep{Hannah2013}. This means that in general the vertical errors in the DEM maps are highest where there are low counts in the SDO/AIA images. This can be due to regions with generally low EM or at the highest temperatures ($>20$ MK), because the sensitivity of SDO/AIA is low and there is usually less plasma emitting at these temperatures. In contrast, the DEM uncertainty is generally low in bright (e.g. flaring) regions \citep{Hannah2013}.

Horizontal errors are derived from the spread of the resolution matrix for each temperature bin and correlate with the quality of the regularized solution in each bin. They indicate whether the emission has been successfully confined to a particular temperature bin (small error) or whether the emission can be attributed to neighboring temperature bins (large error). Large errors are again caused by a weak response and noise in the data \citep{Hannah2012,Hannah2013}.

\begin{figure}
  \resizebox{\hsize}{!}{\includegraphics{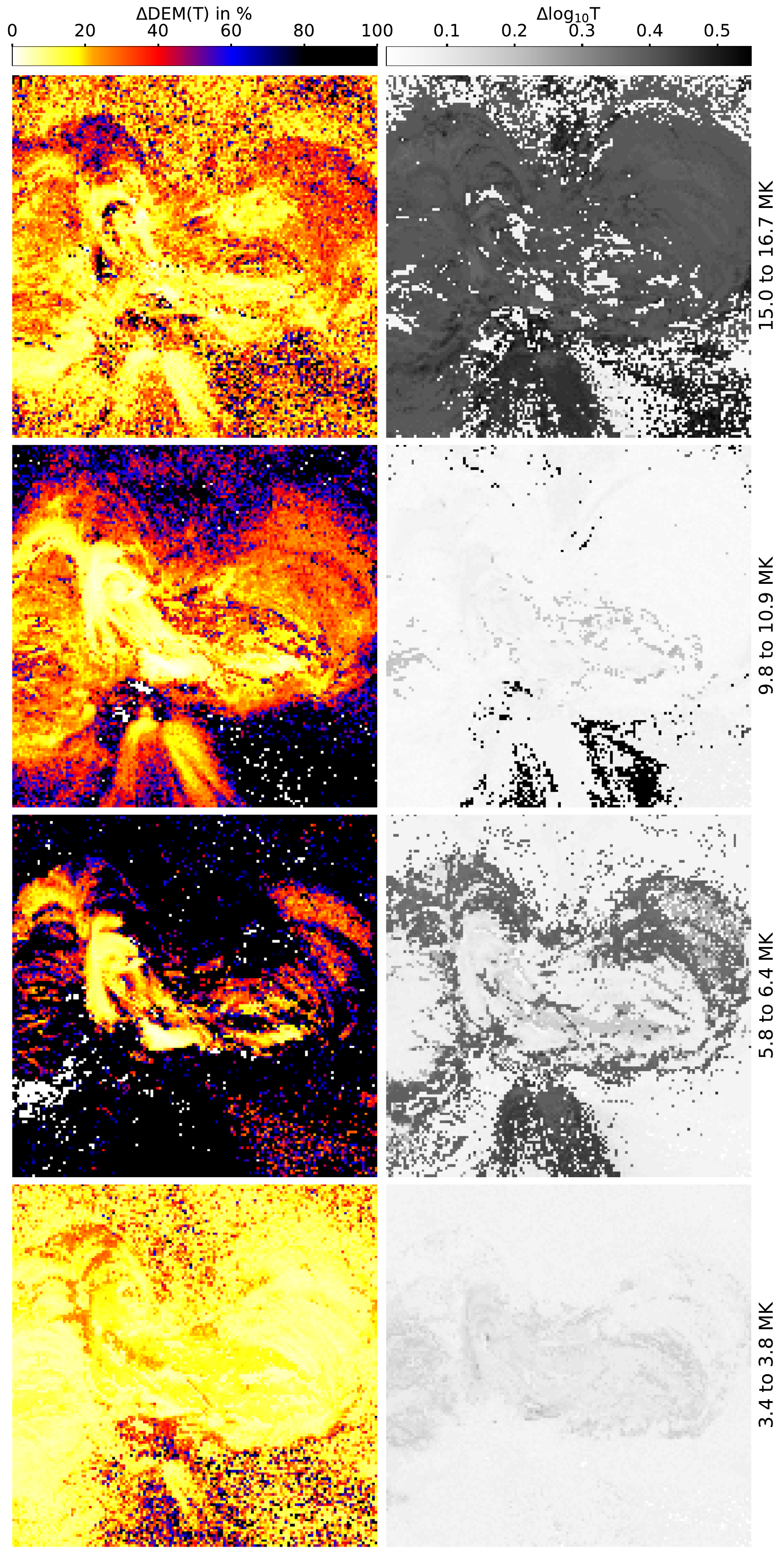}}
  \caption{Maps showing both errors associated with the DEM reconstruction around 09:57:04~UT (see Fig. \ref{fig:DEM_overview}) in four selected temperature bins. Left column: vertical errors ($\Delta \mathrm{DEM(T)}$). Right column: horizontal errors ($\Delta \mathrm{log}_{10} \mathrm{T}$) corresponding to the temperature resolution.}
  \label{fig:DEM_errors}
\end{figure}

To visualize the errors associated with our DEM results (Sect. \ref{sec:results_flare}), we used a plot format from \citet{Hannah2013}, shown in Fig. \ref{fig:DEM_errors}. It shows maps of vertical and horizontal errors for selected temperature bins of the DEM reconstruction performed around 09:57:04~UT (third column in Fig. \ref{fig:DEM_overview}). The selected temperature bins represent the lower and upper bounds of the averaged temperature ranges shown in Fig. \ref{fig:DEM_overview}, giving a good overview of the errors present over the entire temperature range analyzed.

As expected, we find small ($\Delta \mathrm{DEM(T)} \leq 20\% $) vertical errors in the flaring regions throughout all temperature bins. Larger errors are present only in the ambient plasma of the AR surrounding the flare, and are particularly prominent in the 5.8--6.4~MK temperature bin.
At the lower temperatures (3.4--10.9~MK), the horizontal errors (temperature resolution) of the DEM results in the flaring regions are also small ($\Delta \mathrm{log}_{10} \mathrm{T} \leq 0.2$), but they increase for even higher temperatures, reaching values of $\Delta \mathrm{log}_{10} \mathrm{T} \geq 0.3$ in the 15.0--16.7~MK range.

In summary, the error maps show that the reconstructed DEM maps are reliable for discussing the flaring plasma since in these regions the emission and temperature are mostly well-constrained. The only significant errors occur at the upper end of the highest temperature range shown in the top row of Fig. \ref{fig:DEM_overview}. While the reconstructed emission remains accurate, there is a decrease in temperature resolution throughout this temperature interval, with the horizontal error reaching values of $\Delta \mathrm{log}_{10} \mathrm{T} \approx 0.4$ at the upper limit. These errors do not significantly affect the discussion of the results and conclusions presented in Sect. \ref{sec:results_flare}.

\end{appendix}

\end{document}